\documentclass[twocolumn,showpacs,preprintnumbers,amsmath,amssymb,superscriptaddress, floatfix]{revtex4}

\usepackage[normalem]{ulem} 
\usepackage {ulem} 

\usepackage{graphicx}
\usepackage{dcolumn}
\usepackage{bm}

\usepackage{color}

\newcommand{\av}[1]{\left\langle{#1}\right\rangle}
\newcommand{\be}{\begin{equation}}
\newcommand{\bear}{\begin{eqnarray}}
\newcommand{\ee}{\end{equation}}
\newcommand{\ear}{\end{eqnarray}}
\newcommand{\e}{{\rm e}}

\renewcommand{\d}{{\rm d}}

\newcommand{\bel}[1]{\be\label{#1}}
\newcommand{\bearl}[1]{\bear\label{#1}}
\newcommand{\req}[1]{(\ref{#1})}

\begin{document}

\preprint{}

\title{Orientational order and glassy states in networks of
  semiflexible polymers}

\author{Martin Kiemes}%
\affiliation{Institut f\"ur Theoretische Physik, Georg-August-Universit\"at G\"ottingen, Friedrich-Hund-Platz 1, D-37077
  G\"ottingen, Germany} \affiliation{ Max-Planck-Institut f\"ur  Dynamik und
  Selbstorganisations, Bunsenstra{\ss}e 10, D-37073 G\"ottingen, Germany }%

\author{Panayotis Benetatos} \affiliation{Institut f\"ur Theoretische Physik, Georg-August-Universit\"at G\"ottingen, Friedrich-Hund-Platz 1, D-37077
  G\"ottingen, Germany}  \affiliation{Theory of Condensed Matter Group,
  Cavendish Laboratory, University of Cambridge, 19 J. J. Thomson Avenue,
  Cambridge, CB3 0HE, United Kingdom}%

\author{Annette Zippelius}%
\affiliation{Institut f\"ur Theoretische Physik, Georg-August-Universit\"at G\"ottingen, Friedrich-Hund-Platz 1, D-37077
  G\"ottingen, Germany}  \affiliation{ Max-Planck-Institut f\"ur  Dynamik und
  Selbstorganisations, Bunsenstra{\ss}e 10, D-37073 G\"ottingen, Germany }%

\date{\today}

\begin{abstract}
  Motivated by the structure of networks of cross-linked cytoskeletal
  biopolymers, we study orientationally ordered phases in
  two-dimensional networks of randomly cross-linked semiflexible
  polymers. We consider permanent cross-links which prescribe a finite
  angle and treat them as quenched disorder in a semi-microscopic
  replica field theory. Starting from a fluid of un-cross-linked
  polymers and small polymer clusters (sol) and increasing the
  cross-link density, a continuous gelation transition occurs. In the
  resulting gel, the semiflexible chains either display long range
  orientational order or are frozen in random directions depending on
  the value of the crossing angle, the crosslink concentration, and the
  stiffness of the polymers. A crossing angle $\theta\sim 2\pi/M$
  leads to long range $M$-fold orientational order, e.g., ``hexatic'' or
  ``tetratic'' for $\theta=60^{\circ}$ or $90^{\circ}$, respectively.
  The transition to the orientationally ordered state is discontinuous
  and the critical cross-link density, which is higher than that of
  the gelation transition,
  depends on the bending stiffness of the polymers and the cross-link
  angle: the higher the stiffness and the lower $M$, the lower the
  critical number of cross-links. In between the sol and the long
  range ordered state, we always expect a gel which is a statistically
  isotropic amorphous solid (SIAS) with random positional and random
  orientational localization of the participating polymers.

\end{abstract}

\maketitle


\section{Introduction\label{sec:Introduction}}
The cytoskeleton is a network of linked protein fibers which plays an
important role in several functions of eukaryotic cells such as 
maintenance of morphology, mechanics and intracellular transport
\cite{Howard}. Cytoskeletal fibers, such as F-actin can be described
as semiflexible polymers with a behavior intermediate between the two
extreme cases of rigid rods and random coils. The function of the
actin cytoskeleton is modulated by a large number of actin-binding
proteins (ABPs) \cite{Winder,Janmey}. The organization of actin
filaments into networks is regulated by ABPs which can be classified
into two broad categories; {\it cross-linkers} promote binding of the
filaments at finite crossing angles whereas {\it bundlers} promote
formation of bundles consisting of parallel or antiparallel
filaments. F-actin is a polar semiflexible polymer and some ABPs bind
filaments with a specific polarity whereas others are not affected by
the filament's polarity. In order to unravel the physics of these
complex aggregates, {\it in vitro} solutions of actin filaments with
controlled cross-linkers have been studied \cite{Kroy_Bausch}.

The stiffness of semiflexible filaments gives rise to orientational correlations
and promotes the formation of structures with long range orientational
order. The isotropic-nematic transition in solutions of partially
flexible macromolecules has been studied theoretically using the
Onsager approach \cite{Khokhlov_Semenov} or the Maier-Saupe approach
\cite{Warner,Spakowitz}. In Ref. \cite{Spakowitz}, the role of the
solvent is taken into account and a very rich phase transition
kinetics is predicted. Aggregation and orientational ordering of
Lennard-Jones macromolecules with bending and torsional rigidity, with
or without solvent, has
recently been investigated in Ref. \cite{Ma_Hu} using molecular
dynamics simulations. Experimental investigations of the
isotropic-nematic transition in lyotropic F-actin solutions have been
carried out in Ref. \cite{Tang1} and measurements of the associated
order parameter are presented in \cite{Tang2}.

The excluded volume effect is not the only mechanism which drives an
isotropic-nematic transition in solutions of semiflexible
polymers. Assemblies of cytoskeletal filaments exhibit a structural
polymorphism due to interactions mediated by a wide range of ABPs, as
shown in Ref. \cite{Lieleg}. 
Theoretical attempts to study the formation of ordered
structures in this kind of systems involve a generalized Onsager
approach \cite{Liu_Bor,Bor_Bru,Bru}, a Flory-Huggins theory \cite{Zilman_Safran},
and a semi-microscopic replica field theory \cite{PRL}. In
Refs. \cite{Liu_Bor,Bor_Bru,Bru,Zilman_Safran} the filaments are modeled as
perfectly rigid
rods whereas in Ref. \cite{PRL} they are considered to be semiflexible
polymers and the thermal bending fluctuations (finite persistence
length) are fully taken into account.

{\it In vitro} structural studies by Wong {\it et al.} \cite{Wong_Li_Saf}
of F-actin in the
presence of counterions have shown the formation of sheets (``rafts'')
with tetratic order without any cross-linking proteins. It appears that electrostatic interactions are the
main mechanism behind the effective ``$\pi/2$'' cross-linking of the
actin filaments in this experiment.  A theory for tetratic raft
formation by rigid rods and reversible sliding ``$\pi/2$''
cross-linkers has been proposed by Borukhov and Bruinsma
\cite{Bor_Bru}.

In Ref. \cite{PRL}, a three-dimensional melt of identical,
fixed-contour-length semiflexible chains is considered and random
permanent cross-links are introduced. The cross-links are such that they
fix the relative positions of the corresponding polymer segments and
constrain their orientations to be parallel or antiparallel to each
other. The aim of the present study is to examine the effect of
cross-links which prescribe a finite crossing angle on a
two-dimensional version of the previous model. We distinguish the case
of sensitive cross-links that are perceptive of the polymers' polarity and unsensitive cross-links that are not. In contrast to previous studies where the rigid rods have a finite width
which allows for nematic ordering \`a la Onsager, our semiflexible
polymers are considered one-dimensional objects and the sole cause for
the emergence of orientationally ordered phases is the interplay of
the finite persistence length of the polymers and the cross-link
geometry. 

Using the polymer stiffness and the cross-link density as control
parameters, we obtain a phase diagram which involves a sol and various
types of orientationally ordered gels. For appropriate values of the
control parameters and crossing angle $\theta \sim 2\pi/M\, ; M \in \mathbb Z$, we predict
the emergence of an exotic gel with random positional localization and
$M$-fold orientational order (e.g., hexatic for $M=6$ or tetratic for
$M=4$).  Similar phases have been
predicted \cite{Bru} for a different system in three space dimensions:
a semi-dilute solution of charged rods with finite diameter in the
presence of polyvalent ions that have the function of (non-permanent)
cross-linkers and may favor various crossing angles. 
Besides the orientationally ordered phase, we also predict a
statistically isotropic amorphous solid (SIAS) with random positional
and orientational localization of its constituent polymers appearing
right at the gelation transition.

This article is organized as follows. In Section II, we introduce our
model and the Deam-Edwards distribution which parametrizes the
quenched disorder associated with the cross-links. In Section III, the disorder-averaged free energy is presented as a functional of a
coarse-grained field which plays the role of an order parameter. Then,
in Section IV, we discuss different variational Ans\"atze that express
positional and orientational localization. The symmetries imposed by
the cross-linking constraints allow the emergence of specific
orientationally ordered phases. The corresponding free energies are
calculated variationally in the saddle-point approximation and a phase
diagram is obtained. We summarize and give an outlook in Section
V. Finally, details of the calculations are given in the appendices.


\section{Model\label{sec:Model}}

We consider a large rectangular two-dimensional volume (area) $V$
which contains $N$ identical semiflexible polymers. A single polymer
of total contour length $L$ is represented by a curve in $2d$ space
with $\mathbf r(s)$ denoting the position vector at arc-length
$s$. Bending a polymer costs energy according to the effective free
energy functional (``Hamiltonian'') for the wormlike chain (WLC) model
\cite{Saito1967}
\begin{equation}
\label{equ:WLC} 
{\cal H}_{WLC}(\{\mathbf r(s)\}) = \frac12 \kappa \int_0^L \d
s \left(\frac {\d \mathbf t(s)}{\d s}\right)^2. 
\end{equation} 
Here we have introduced the tangent vector $\mathbf t(s)=\d \mathbf
r(s) / \d s$ and chosen a parametrization of the curve, in accord with
the local inextensibility constraint of the WLC, such that $|\mathbf
t(s)| = 1$.  The position vector is recovered from $\mathbf t(s)$ as
$\mathbf r(s)= \mathbf r(0) +\int_0^s \mathbf t(\tau) \d\tau$. Hence
the conformations of a single polymer, that can be bent but not be
stretched, are alternatively characterized by $\mathbf t(\tau),\; 0\leq \tau
\leq L$, and $\mathbf r(0)$. The bending stiffness is denoted by
$\kappa$; it determines the persistence length $L_p$ according to
$\kappa = L_p k_B T/2$. Throughout the rest of the paper we set $k_B T
= 1$. The WLC model can describe stiff rods, obtained in the limit
$L/L_p\to 0$ and fully flexible coils, obtained in the limit $L_p/L\to
0$.

The mutual repulsion of all monomers is modeled by the excluded-volume
interaction
\begin{eqnarray} \label{equ:H_eV}
{\cal H}_{ev} &=& \frac{N^2}{2 V}  \sum_{\mathbf k \neq 0} \sum_{m \in \mathbb Z}   \lambda^2_{|\mathbf k|, m}|\rho_{\mathbf k, m}|^2\;
\end{eqnarray}
with 
\begin{equation}
\rho_{\mathbf k, m} := \frac 1N \sum_{i=1}^N \frac 1L \int_0^L \d s\;  \e^{i \mathbf k \mathbf r_i(s) } \; \e^{i m \psi_i(s) }
\end{equation}
being the Fourier transformation of the positional-orientational
density. $\mathbf t_{i}(s)=(\cos{\psi_i(s)},\sin{\psi_i(s)})$ denotes
the orientation of monomer $s$ on polymer $i$. The coefficients
$\lambda_{|\mathbf k|, m}^2$ depend only on the absolute value of the
vector $\mathbf k$ in order to preserve the rotational symmetry of the
system. They are later chosen large enough to provide stability with
respect to density modulations. For details, see Appendix
\ref{sec:1-RS}.

The Hamiltonian ${\cal H}_{WLC}$ is invariant with respect to
interchanging head and tail of the filaments, i.e., the energy is
unchanged under reparametrizations of the contour of one polymer $i$
by $ \{ \mathbf r_i(s) \to \mathbf r_i(L-s), \forall s \in
[0,L]\}$. However, in the following, we want to consider filaments
with a definite polarity which could for example arise due to the
helicity of F-actin. The WLC Hamiltonian is not sensitive to such a
polarity, but the cross-links may or may not differentiate between the
two states of the filament, as discussed below.

\begin{figure}[h]
 \includegraphics[width=8.5cm]{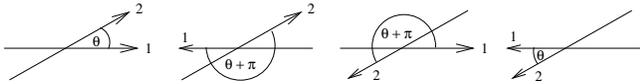} 
 \caption{\label{fig:unpolarX}4 possibilities for two filaments to be
   joined by an unsensitive cross-link.}
\end{figure}

We now introduce $M$ permanent (chemical) cross-links between pairs of
randomly chosen monomers. A single cross-link, say between segments
$s$ and $s'$ on two chains $i$ and $j$, constrains the two polymer
segments to be at the same position, i.e. $\mathbf r_i(s) = \mathbf
r_j(s')$, and fixes their relative orientation by a constraint
expressed by the function $\Delta \left(\mathbf t_i(s), \mathbf
  t_j(s'),\theta\right)$. The constraint of relative orientation is
most easily formulated in polar coordinates where $\Delta$ just fixes
the crossing angle $\psi - {\psi'}$ to some prescribed value $\theta$.
In the case of cross-links {\it sensitive} to polarity we simply have
\bel{equ:x-link} \Delta (\psi - \psi',\theta) \;=\; \delta\big(\psi -
{\psi'} - \theta \big), \ee where the weight of the delta function is
such that $\int_0^{2\pi} \frac{\d \psi}{2\pi}\, \delta(\psi) = 1$.  In
the {\it unsensitive} case on the other hand, changing head and tail
of either filament leads to four equivalent situations that correspond
to two different crossing angles (see Fig.~\ref{fig:unpolarX}).  In
this case the cross-link \req{equ:x-link} thus appears with the two
equally likely cross-linking angles $\theta$ and $\theta + \pi$. We
point out that in our model the polarity is only recognized by the
cross-links and does not enter into the Hamiltonian.

The delta function in \req{equ:x-link} is the simplest way to model
the angular constraint of the cross-links. It is however much more
realistic to introduce an effective angular cross-linking potential
that allows for fluctuations around the preferred direction. A
simple model for these ``soft'' cross-links is given by
\begin{equation}\label{equ:soft_xlink}
 \Delta(\psi - \psi', \theta) = \frac{1}{I_0(\gamma)} \e^{\gamma \cos(\psi-\psi' -\theta)}\qquad\mbox{.}
\end{equation}
The cross-link stiffness parameter $\gamma$ controls the variance
of the fluctuations of the angle and it is clear that in the
limit $\gamma \rightarrow \infty$ we recover the simple delta function
model of ``hard'' cross-links. In the following we will first explore
the case of hard cross-links, because it is mathematically simpler. It
gives rise, however, to artifacts which disappear when considering the
more physical model which favors certain angles but does not enforce
them strictly.

The cross-links are permanent and do not break up or rebuild. Hence we
are led to study the equilibration of the thermal degrees of freedom
$\{\mathbf r_{i}(s)\}$ in the presence of {\it quenched} disorder
represented by a given cross-link configuration 
${\cal C}=\{i_e,j_e;s_e,s'_e\}_{e=1}^M$ which is characterized by the set of pairs of polymer segments that are involved in a cross-link. The central quantity of interest
is the canonical partition function
\begin{equation}
\label{crosslinks}
Z({\cal C})=\av{\prod_{e=1}^M \delta(\mathbf r_{i_e}(s_e) - \mathbf r_{j_e}(s'_e))
\Delta \left(\mathbf t_{i_e}(s_e) - \mathbf
  t_{j_e}(s'_e),\theta\right)}^{\cal H}.
\end{equation}
Here the thermal average $<...>^{\cal H}$ is taken with respect to the Hamiltonian ${\cal H} := {\cal H}_{WLC} + {\cal H}_{ev}$ and the partition function depends on the
cross-link realization ${\cal C}$. The free energy $F$ is expected to
be self-averaging so that we are allowed to compute the
disorder averaged free energy $[F]=- [\ln Z]$, where $[...]$ denotes
the ``quenched'' average over all cross-link configurations according to some
distribution ${\cal P}({\cal C})$.
We assume that the different realizations obey a Deam-Edwards-like \cite{DeamEdwards}
distribution 
\begin{equation} \label{Deam}
 {\cal P}({\cal C}_M) \;\propto\; \frac 1{M!} \left(\frac{\mu^2 V}N\right)^M \av{\prod_{e=1}^M \delta\left(\mathbf r_{i_e}(s_e) - \mathbf r_{j_e}(s_e') \right) }^{\cal H},
\end {equation}
implying that polymer segments that are likely to be
close to each other in the uncross-linked melt also have 
a high probability of getting cross-linked. Note that the probability of
cross-linking is independent of the relative orientation of the two
segments, so that the cross-link actually reorients the two participating
segments. The parameter $\mu^2$ controls the mean number of
cross-links in the system: $[M]/N \; \sim \; O(\mu^2)\;$.

Given the Hamiltonian, the constraints due to cross-linking and the
distribution of 
cross-link realizations, the specification of the model
is complete and we proceed to calculate the disorder averaged free
energy $[F]$.


\section{Replica free energy\label{sec:repl_f+OP}}

The standard method to treat the quenched disorder average is the replica method, representing the disorder averaged free energy as $[F] = - [\ln Z] =- \lim_{n\rightarrow 0} ([Z^n] -1)/n$, i.e., we have to calculate the simultaneous disorder average of the partition sums of $n$ non-interacting copies of our system. In the end, we extract the disorder averaged free energy $[F]$ from $[Z^n]$ as the linear order coefficient of its expansion in the replica number $n$.

In this section, we present only the essential formulas and present  the details of the calculation in Appendix \ref{sec:F_and_SP}. The disorder average, $[Z^n]$, gives rise 
to an effectively uniform theory, however, with a coupling of different replicas: 
\begin{widetext}
\begin{eqnarray}
\label{avZ}
[Z^n]\propto\Bigg< \exp\Bigg( \frac{\mu^2 V}{2N} \sum_{i,j=1}^N
\int_{s,s'}  \delta(\hat {\mathbf r}_{i}(s) - \hat{\mathbf r}_{j}(s') ) \;\Delta(\check \psi_{i}(s) - \check
\psi_{j}(s'), \theta ) \Bigg) \Bigg>^{\cal H}_{n+1}\;,
\end{eqnarray}
\end{widetext}
where the average $\langle...\rangle^{\cal H}_{n+1}$ is over the $(n+1)$-fold replicated Hamiltonian ${\cal H}$. To simplify the notation we have introduced the abbreviations $\hat {r} \equiv \left({\mathbf r}^0,{\mathbf r}^1,\dots,{\mathbf r}^n \right)$,
$\check \psi \equiv \left( \psi^1,\dots,\psi^n \right)$, and the shorthand notation $\int_s \equiv (1/L) \int_0^L \d s$. Furthermore, $\delta(\hat {\mathbf r}) \equiv \prod_{\alpha=0}^n \delta({\mathbf r}^\alpha)$ and $\Delta$ denotes for sensitive cross-links 
\be
\Delta_s(\check \psi, \theta) \equiv \prod_{\alpha=1}^n \delta( \psi^\alpha - \theta)
\ee 
and  
\be 
\Delta_u(\check \psi, \theta) \equiv \frac 12 \left\{ \prod_{\alpha=1}^n \delta(\psi^\alpha - \theta) + \prod_{\alpha=1}^n \delta(\psi^\alpha - (\theta+\pi))\right\}
\ee
if they are unsensitive to polarity of the filaments. For soft cross-links we introduce the corresponding definitions.

Different polymers are decoupled via a Hubbard-Stratonovich transformation,
introducing collective fields, $\Omega(\hat {\mathbf x}, \check \varphi)$, 
whose expectation values are given by
\begin{equation}
\label{equ:op_realspace}
\big \langle \Omega(\hat {\mathbf x}, \check \varphi) \big \rangle
=\frac{1}{N} \sum_{i=1}^N\int_s \big \langle 
\delta(\hat {\mathbf x}- \hat {\mathbf r}_i(s))  \delta(\check \varphi -
\check \psi_i(s)) \big \rangle.
\end{equation}
The field $\av{\Omega}$ quantifies the probability to find monomer $s$
on chain $i$ at
position ${\mathbf x^0}$ in replica $0$, at position ${\mathbf x^1}$
in replica $1$,... and at position ${\mathbf x^n}$
in replica $n$ and similarly to find it oriented in the direction ${\mathbf
  e^1}=(\cos(\varphi^1),\sin(\varphi^1))$ in replica $1$,... and
oriented in the direction ${\mathbf
  e^n}=(\cos(\varphi^n),\sin(\varphi^n))$ in replica $n$.
Sometimes it is convenient to use its equivalent representation in Fourier space
\begin{equation}
\label{equ:op_fourierspace}
\big \langle \Omega(\hat {\mathbf k}, \check m) \big \rangle = 
 \frac 1{N} \sum_{i=1}^N \int_s \big \langle \e^{i \hat {\mathbf k}
  \hat {\mathbf r}_i(s)} \;
 \e^{i \check m \check \psi_i(s) }\big \rangle .  
\end{equation}

In terms of these collective fields the effective replica theory is given by
\begin{equation} \label{equ:field_theory}
  [Z^n] \;\sim\; \int {\cal D} \{\Omega\} \exp\left(- N {\cal F} (\{\Omega\}) \right). 
\end{equation}
The replica free energy per polymer reads
\begin{eqnarray}\label{equ:replica_F}
{\cal F}=\frac{\mu^2 }{2 V^n} \overline \sum_{\hat {\mathbf k}} \sum_{ \check m} \Delta   |\Omega|^2   - \ln z(\{\Omega\})\;,
\end{eqnarray}
with the effective single chain partition function
\begin{widetext}
\begin{eqnarray}
\label{single_chain}
 z(\{\Omega\}) =
\Bigg\langle \exp \bigg(  \frac {\mu^2}{2 V^n} 
\overline \sum_{\hat {\mathbf k}}\sum_{\check m} \Delta\; \Omega  \; \int_s \exp\Big\{- i \hat {\mathbf k} \hat {\mathbf r}(s)\Big\}  \exp\Big\{- i \check m \check \psi(s) \Big\} \bigg) \Bigg\rangle^{{\cal H}_{WLC}}_{n+1}\quad .
\end{eqnarray}
\end{widetext}
The average $\langle...\rangle_{n+1}^{WLC}$ refers to the
$(n+1)$-fold replicated Hamiltonian of a single wormlike chain.
The sum $\overline \sum_{\hat {\mathbf k}}$ over replicated wave vectors is restricted to the combination of the so-called ``0-replica sector'' (0RS) that contains only the point $\hat
 {\mathbf k} = (0,\dots,0)$ and the ``higher-replica sector'' (HRS) where at least
two wave vectors in different replicas are non-zero: ${\mathbf
  k}^{\alpha}\neq 0$ and ${\mathbf k}^{\beta}\neq 0$ with $\alpha \neq
\beta$.

Note that in the above overview we left out the ``1-replica sector'' (1RS) that 
consists of $(n+1)$-fold replicated vectors $\hat {\mathbf k}^\alpha$ where only 
one entry (the $\alpha$th) is non-zero, i.e. $\hat {\mathbf k}^\alpha =
(0,...,{\mathbf k}^{\alpha},...,0)$, and $\check m$ being arbitrary. The corresponding fields $\tilde
\Omega^\alpha({\mathbf k}, \check m)$ describe regular macroscopic density 
fluctuations (modulated states). In this work we focus on the properties of 
the macroscopically translationally invariant amorphous solid state and assume that the inter-polymer interactions \req{equ:H_eV} are such that periodic density fluctuations are suppressed. See Appendices \ref{sec:F_and_SP} 
 and \ref{sec:1-RS} for more details on that issue.


\section{Variational approach}\label{sec:variat_appr}

We shall only discuss the saddle-point approximation to
the field theory on the right hand side of (\ref{equ:field_theory}) replacing
the field $\Omega$ by its saddle point $\Omega_{sp}$ which has to be
calculated from $\delta {\cal F}/\delta \Omega|_{\Omega_{sp}} = 0$. 
In fact, even the saddle point equation is too hard to solve, because the 
conformational distribution of the WLC is very complex. Similarly, it is not possible to
perform a complete stability analysis of the Gaussian theory.
Hence we restrict ourselves to a variational approach and in the
following we are going to construct Ans\"atze which capture the symmetry
of the  different physical states which may emerge.	

What behavior do we expect?
Upon increasing the number of cross-links up to about one per polymer, there should be a gelation transition from a liquid to an amorphous solid state. A finite fraction of the polymers form the percolating cluster and are localized at random positions. The other polymers belong to finite clusters or remain un-cross-linked and are still free to move around in the volume. This scenario has been found to be valid for a variety of different models \cite{p_beine,Castillo96, Theissen, Ulrich06, dirpols}. A similar replica field-theoretic approach has beeen used by Panyukov and Rabin to study the properties of well cross-linked macromolecular networks \cite{Rabin1,Rabin2}.

For semiflexible polymers the positional localization in the macroscopic cluster is accompanied by orientational localization. The cross-links create locally an orientational structure. Supposing that the semiflexible polymers are rather stiff, a long range orientationally ordered state can be established. Otherwise we may find an orientational glass, i.e., the directions of the polymer segments are frozen in random directions in analogy to the low temperature phase of a spin glass\cite{spin_glass_theory}. We call such a phase statistically isotropic amorphous solid (SIAS).

\begin{figure}
\begin{center}
 \includegraphics[width=6cm]{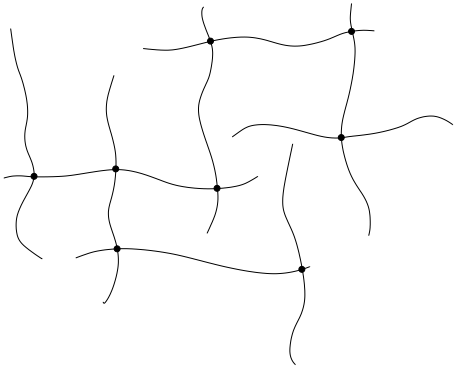}
\caption{\label{fig:4fold}Sketch of the tetratic phase (M=4).}
\end{center}
\end{figure}

\begin{figure}
\begin{center}
 \includegraphics[width=6cm]{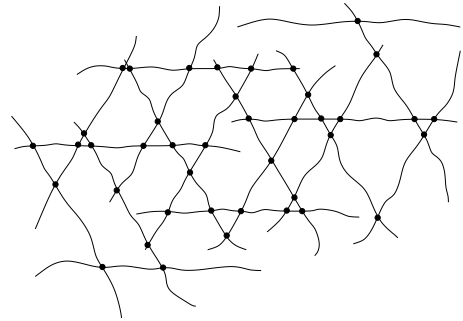} 
\caption{\label{fig:3/6fold}Sketch of the triangular phase (M=3) for sensitive or the hexatic phase (M=6) for unsensitive cross-links.}
\end{center}
\end{figure}

Can we expect long range orientational order for arbitrary  crossing angles $\theta$? 
We first consider a special case, namely that the crossing angle is such that an integer multiple of the crossing angle, $M \theta$, equals $2\pi$, e.g. $\theta=120^{\circ}$ and $M=3$. Such a choice of crossing angle allows for orientationally ordered states provided the chains are sufficiently stiff. In the case of sensitive cross-links we expect $M$-fold discrete rotational symmetry, in the case of unsensitive cross-links and odd M we expect $2M$-fold symmetry because cross-links are established including both the angles $\theta$ and $\theta + \pi$ (see Fig. \ref{fig:unpolarX}). Sketches of gels with long range four-fold order ($\theta=90^{\circ}$) or long range three-fold or six-fold order respectively ($\theta=120^{\circ}\,/\, 60^\circ$) are shown in Fig.~\ref{fig:4fold} and Fig.~\ref{fig:3/6fold}.
The case $\theta = k \frac{2 \pi}M$ with $k=1,\dots, M-1$  leads in a
similar fashion to an M-fold symmetric phase.

Let us now consider the constraints on the order parameter imposed by symmetries. To this end we note that the Hamiltonian (\ref{equ:WLC}), the cross-link constraints (\ref{crosslinks}) and the Deam-Edwards distribution (\ref{Deam}) are invariant with respect to uniform translations and rotations of all particles positions $\{\mathbf r_{i}(s)\}$. The disorder averaged partition function of Eq.(\ref{avZ}) is invariant under spatially uniform translations and spatially uniform rotations of {\it each replica separately}.

Only the fluid state is expected to have the full symmetry of the partition function Eq.(\ref{avZ}) because here, the polymers and finite clusters of polymers are free to sample the complete volume and can take any orientation. This implies for the order parameter
\begin{equation}
\big \langle \Omega(\hat k, \check m) \big \rangle =
\delta_{\hat k,\hat 0}\delta_{\check m,\check 0}\;,
\end{equation}
i.e., each replica is separately invariant under translations
and rotations.

We will only investigate gels with {\it random} localization of a fraction of the particles. In other words, we restrict ourselves to amorphous solids and do not consider periodic density fluctuations. It might be of interest to study one-dimensional periodic density modulations with an overall orientation of the WLCs perpendicular to the wave vector,-- reminiscent of bundles. However this is not the topic of the present paper, where we stick to the incompressible limit. 

If the translational and rotational symmetry is broken 
spontaneously, then we expect a non-trivial expectation value of the
local density
\begin{equation}
 \rho_{i,s}(\mathbf x,\varphi):= \delta({\mathbf x} - {\mathbf r}_i(s) ) \; 
  \delta(\varphi - \psi_i(s) ) 
\end{equation}
in a particular equilibrium state. 
For example, in the {\it amorphous solid} phase a finite fraction of
particles should be spatially localized at preferred positions 
and possibly oriented along preferred directions. A simple Ansatz 
quantifying such a scenario is the following: 
\begin{equation}
  \langle \rho_{i,s}(\mathbf x,\varphi) \rangle \propto 
  \e^{-\frac 1{2 \xi^2}(\mathbf x - \mathbf a)^2} 
\e^{\eta \cos(\varphi - \varphi_0)}.
\label{equ:ansatz_single_system_polar}
\end{equation} 
The vector ${\mathbf a}$ is the preferred mean position of monomer $s$ and the thermal fluctuations
around the preferred position are controlled by the localization length $\xi$. The preferred orientation is given by $\varphi_0$ and $\eta$ parametrizes the variance of the orientational distribution.\\

An amorphous solid is {\it macroscopically translational invariant}, so that
the spontaneous symmetry breaking of
translational invariance is {\it local}. All macroscopic
observables should display translational symmetry and in particular all
moments of the local density 
\begin{eqnarray}
 & \tilde{\Omega}^J({\mathbf k}_1,... {\mathbf
   k}_J,m_1,... m_J)  \\ & 
: =\frac{1}{N} \sum_{i=1}^N\int_s  \langle  \rho_{i,s}(\mathbf k_1,m_1)
\rangle ... \langle \rho_{i,s}(\mathbf k_J,m_J)
 \rangle\nonumber\\
 & \sim  \delta_{{\mathbf k}_1+...{\mathbf k}_J,{\mathbf 0}} \nonumber
\end{eqnarray}
are non-zero only, if the wave vectors add up to zero (see reference \cite{Castillo96} for a more detailed discussion).\\

As far as the rotational symmetry is concerned, we will consider two
possibilities: a statistically isotropic state (SIAS)
 as well as states with true long range orientational order. In
the former case the spontaneous symmetry breaking of
rotational invariance is {\it local} and restored globally analogously to what happens in the case of translational symmetry. All macroscopic properties, such as the moments, are non-zero only, if
the angular momentum ``quantum'' numbers add up to zero:
\begin{equation}
   \tilde{\Omega}^J({\mathbf k}_1,... {\mathbf
   k}_J,m_1,... m_J)  \sim  \delta_{{\mathbf k}_1+...{\mathbf
     k}_J,{\mathbf 0}}\,
\delta_{m_1+...m_J,0}
\end{equation}

For the long range ordered case the simplest Ansatz consists in assuming a local $M$-fold symmetry. This implies for the moments 
\begin{equation}
   \tilde{\Omega}^J({\mathbf k}_1,... {\mathbf
   k}_J,m_1,... m_J)  \sim  \delta_{{\mathbf k}_1+...{\mathbf k}_J,{\mathbf 0}}
 \, \prod_{\alpha=1}^J \delta_{m_\alpha, {\mathbb Z} M}\;,
\end{equation} 
i.e. each $m_\alpha$ has to equal an integer multiple of $M$.

Note that a rotation affects both $\mathbf r_i(s)$ and $\mathbf
t_i(s)$ and that consequently, a system with e.g. $M$-fold symmetry
should be described by an Ansatz $\Omega(\hat {\mathbf k}, \check
\varphi)$ which is only invariant under common rotations of the $\mathbf
k^\alpha$ and $\varphi^\alpha$. We have chosen our approach for
simplicity and incorporated the symmetry with respect 
to the orientational and position vector separately. Consequently, the
Ansatz is symmetric under individual rotations of the spatial and
angular argument. An interesting extension of the present work would involve the study of Ans\"atze which couple positional and orientational degrees of freedom. Some preliminary work in this direction has been done in the context  of random networks of covalently connected atoms or molecules\cite{Shakhnovich1999, Goldbart2002}.\\

Rephrasing our results in replica language, frozen fluctuations in a single equilibrium state
correspond to a non-zero expectation value of the density,
$\rho_{i,s}({\mathbf k},m)$, within one
replica. The full statistics of the local static fluctuations that we need for the characterization of the macroscopic symmetries in the gel is specified by the order parameter field $ \Omega(\hat {\mathbf k}, \check m)$, encoding
moments of arbitrary order. In particular macroscopic translational invariance
requires $\sum_{\alpha=0}^n\mathbf k_{\alpha}=0$; macroscopic
rotational invariance requires $\sum_{\alpha=1}^nm_{\alpha}=0$; long
range orientational order for crossing angle $\theta=\frac{2\pi}{M}$
requires $ m_{\alpha}=l M$ with $l \in {\mathbb Z}$.\\

Let us now encode these physical expectations in a variational Ansatz:
The liquid state is characterized by $\Omega_{sp}(\hat {\mathbf k}, \check m) = \delta_{\hat{\mathbf k},\hat 0} \delta_{\check m,\check 0}$. It becomes unstable at a critical cross-link density, $\mu^2_c$, where a percolation transition takes place and a macroscopic cluster is formed containing a finite fraction of the polymers. To account for the fraction of localized particles $Q$ in the percolating cluster coexisting with mobile particles (fraction $1-Q$) in finite clusters, 
we make the following general Ansatz for the expectation value of the order
parameter:
\begin{equation} 
\Omega_{sp}(\hat{\mathbf k}, \check \varphi) = (1-Q) \; \delta_{\hat{\mathbf k},\hat {\mathbf 0}}  + Q \; \omega(\hat{\mathbf  k}^2, \check \varphi)\; \delta_{{\mathbf 0},\sum_{\alpha=0}^n {\mathbf k}^\alpha} \label{equ:generic_order_parameter} 
\end{equation}
The first term describes the sol phase which is characterized by perfect spatial homogeneity and orientational isotropy. The second part describes the amorphous solid which is macroscopically translational invariant as reflected in $\delta_{{\mathbf 0},\sum_{\alpha=0}^n {\mathbf k}^\alpha} $. The details of the amorphous solid
phase under consideration have to be implemented in the order
parameter $\omega(\hat{\mathbf k}^2, \check\varphi)$.

Plugging the generic form of the order parameter \req{equ:generic_order_parameter} into the replica free energy \req{equ:replica_F} we get 
\begin{widetext}
\begin{eqnarray} 
&& {\cal F}_{sp}= 
  \frac{\mu^2}{2 V^n}\Big\{ 1-Q^2 +  Q^2 \;
 \overline \sum_{\hat {\mathbf k}}\int_{\check \varphi \check \varphi'}  
  \omega(\hat{\mathbf k}^2, \check \varphi) \;\Delta(\check \varphi- 
  \check \varphi',\theta)\; \omega(\hat{\mathbf k}^2, \check \varphi')
  \delta_{{\mathbf 0}, \sum_{\alpha=0}^n \mathbf k^\alpha}\Big\} 
\label{equ:replica_F_gen_op}\\ 
 &-&\frac{\mu^2}{V^n} (1-Q) 
  - \ln \Bigg<\exp\Big\{\frac{\mu^2 Q}{V^n} \; 
\overline \sum_{\hat{\mathbf k}}\int_{\check \varphi \check \varphi'}  
  \delta_{{\mathbf 0}, \sum_{\alpha=0}^n \mathbf k^\alpha}\; \Delta(\check \varphi - 
  \check \varphi', \theta) 
\;\omega(\hat{\mathbf k}^2, \check \varphi) 
  \int_s \; \e^{i\hat {\mathbf k} \hat {\mathbf
      r}(s)} 
\delta(\check\varphi' - 
\check \varphi(s)) \Big\}\Bigg>_{n+1}^{{\cal H}_{WLC}}
\nonumber 
  \end{eqnarray}
\end{widetext}
We expect the gel fraction to be small close to the gelation transition and thus
expand the log-trace contribution of the free energy 
in $Q$. The terms linear in $Q$ cancel, as they should for the
expansion around $Q = 0$ to be justified.


\subsection{$M$-fold orientationally ordered amorphous  solid \label{sec:m-fold}}
\subsubsection{Hard cross-links}
Assuming that the thermal fluctuations of the polymers are to a
certain degree suppressed by their stiffness and a sufficient number 
of cross-links has been formed, long range orientational order 
may be present as sketched in Fig.~\ref{fig:4fold} and
Fig.~\ref{fig:3/6fold}. The Ansatz for the $M$-fold
orientationally ordered amorphous solid favors $M$
preferred orientational axes separated by angles $\frac{2\pi}{M}$.
A simple way to incorporate this symmetry is the following Ansatz for
the replica order parameter 
\begin{equation} 
 \omega ( \hat {\mathbf k}, \check
\varphi ) = \e^{-\frac{\xi^2}2 \hat {\mathbf k}^2} \; \frac{
  \e^{\eta \sum_{\alpha=1}^n \cos(M \varphi^\alpha)}}
{I_0^n(\eta)}. \label{equ:Ansatz_mfold}
\end{equation}
Here $I_0$ denotes the modified Bessel function of the first kind; it ensures the proper normalization.

For $\hat {\mathbf  k} = \hat {\mathbf 0}$, the above order parameter is a probability distribution and thus specifies 
 the local orientational order completely. In experiment on the other hand one
has access to low order moments only. The simplest physical order parameter being
sensitive to the degree of long range $M$-fold orientational order is
given by
\bearl{equ:mfold_OP}
{\cal S}_M &:=&\frac{1}{N} \sum_{i=1}^N  \int_s \;\Big\langle \cos(M \psi_i(s))\Big\rangle \\
&\sim& \Big[\Big\langle \cos(M \psi_i(s))\Big\rangle \Big]  \label{equ:mfold_OP_SA}\\
&=& \frac{\eta}{2} + {\cal O}(\eta^2) \nonumber 
\ear
where in the second line, we replaced the average over all monomers of the system by the disorder average of one arbitrary monomer. This expression allows to relate ${\cal S}_M$ to our effective one-particle theory. $\av{\dots}$ denotes the thermal expectation value of the system for a given instance of disorder ${\cal C}$.
In order to establish the connection between the variational parameter $\eta$ and the physical order 
parameter ${\cal S}_M$ we evaluate \req{equ:mfold_OP_SA} using \req{equ:Ansatz_mfold}
and find that ${\cal S}_M$ is in leading order proportional to $\eta$.

We plug the Ansatz \req{equ:Ansatz_mfold} in \req{equ:replica_F_gen_op} and
perform the $\hat {\mathbf k}$ summations and $\check \varphi$
integrations. If we have chosen $M$ according to the symmetry considerations of the previous section the constraint $\Delta$ drops out. We are left with the
free energy as a function of three variational parameters: the gel fraction $Q$,
the spatial localization length $\xi$ and the degree of orientational
order as measured by $\eta$. To determine these parameters we minimize
the free energy. The resulting equation for the gel fraction is
universal and has been derived previously~\cite{Castillo96}. There are
two solutions $Q = 0$ and $Q \sim \frac 2{\mu^4} (\mu^2 - 1)$ which implies that the phase
transition from sol to gel takes place at $\mu_c^2 = 1$.

In order to determine the remaining parameters $\frac 1 {\xi^2}$ and $\eta$ we assume them to be small near the transition and do a Taylor expansion of the free energy where we leave out terms which do not depend on $\xi$ or $\eta$ because these terms are irrelevant for the variation of ${\cal F}$. More precisely, it will turn out that $\frac{1}{\xi^2} \propto Q $, being thus small close to the transition, and we will keep contributions up to order $1/\xi^2$ in our expansion. As for the variational parameter $\eta$, we will find that it actually jumps from $0$ to a finite value, i.e that there is a first order orientational transition. We expand all the same in $\eta$ and obtain at least a qualitative picture. Details on the calculations are given in Appendix \ref{sec:m-fold_calc}. 

Since our Ansatz is replica-symmetric it is straightforward to extract the part of ${\cal F}$ linear in the replica index $n$ and we find
\begin{widetext}
  \begin{eqnarray}\label{equ:var_free_energy_mfold} 
\frac{{\cal F}}{n}&=& \frac{\mu^2 Q^2}{2} \Bigg\{  - \frac{ \mu^4 Q}{6} \ln \frac{L^2}{\xi^2} + \frac{\mu^2 L^2}{4 \xi^2} \; g(\frac{L}{L_p})  + \delta_{M,1} \frac{\mu^2 L^2}{4 \xi^2} \Big\{ \eta^2 l(\frac{L}{L_p}) - \frac{\eta^4}{16} \tilde l(\frac{L}{L_p}) + \frac{\eta^6}{16} \tilde{\tilde l}(\frac{L}{L_p})\Big\} \\
& &\hspace{1.2cm}+ \frac{\eta^2}{2}\Big(1- \mu^2 \; h(\frac{M^2 L}{L_p}) \Big) - \frac{7}{32} \eta^4 \Big(1 - \mu^2 \; 
\tilde h(\frac{M^2L}{L_p}) \Big) + \frac{31}{288} \eta^6 \Big(1 - \mu^2 \;\tilde {\tilde h}(\frac{M^2L}{L_p}) \Big)\;\;\Bigg\} \nonumber
  \end{eqnarray}
\end{widetext} 
Note that for $M>1$ there are no terms coupling the orientational order specified by $\eta$ and the positional localization specified by $\xi^2$. Such terms appear in higher order but will not be considered here because we restrict ourselves to the vicinity of the gel point. The persistance length $L_p$ and the localization length $\xi$ are both rescaled by the contour length $L$ of the polymers.

The functions $g$, $h$, $\tilde h$, $\tilde {\tilde h}$  and $l$,  $\tilde l$, $\tilde {\tilde l}$ go to zero for large argument and are for small argument approximately given by
\begin{eqnarray} 
&& g(x)\;\sim\; \frac{1}{6}-\frac{1}{30} x \;\;\;\;\mbox{,}\;\;\;\;
 h(x)\;\sim\;1-\frac{1}{3}x \nonumber \\
&&\tilde h(x) \;\sim\; 
1 - \frac 8{21} x  \;\;\;\;\mbox{,}\;\;\;\;    \tilde {\tilde h}(x)
\;\sim\; 1 - \frac {48}{93} x \nonumber \\
&&l(x)\;\sim\; \frac{1}{6}-\frac{1}{10} x \;\;\;\;\mbox{,}\;\;\;\;
\tilde l(x)\;\sim\;\frac 13 - \frac15 x  \nonumber \\
&& \tilde {\tilde l}(x) \;\sim\;\frac{125}{72}-\frac{61}{30} x
\label{gofx_hofx}
\end{eqnarray}
For the definitions of these functions see Appendix \ref{sec:exp_val_M-fold}.

Minimizing ${\cal F}$ for $M>1$ with respect to $\xi^2$ yields
\begin{equation}
\frac{1}{\xi^2}= \frac 2{3 L^2} \frac{\mu^2 Q}{g(\frac L{L_p}) } \,=\, \frac {\mu^2 Q}{3 R_g^2}\hspace{1cm}\mbox{.}
\end{equation}
Hence the filaments are localized as soon as a percolating cluster of
cross-linked chains has formed and the gel fraction is finite. The
localization length is independent of $M$ and its scale is set by the
radius of gyration $R_g$ of the filament which is $\sim L^2$ for stiff
chains and $\sim L$ for random coils.

\begin{figure}
\begin{center}
\includegraphics[width=7.6cm]{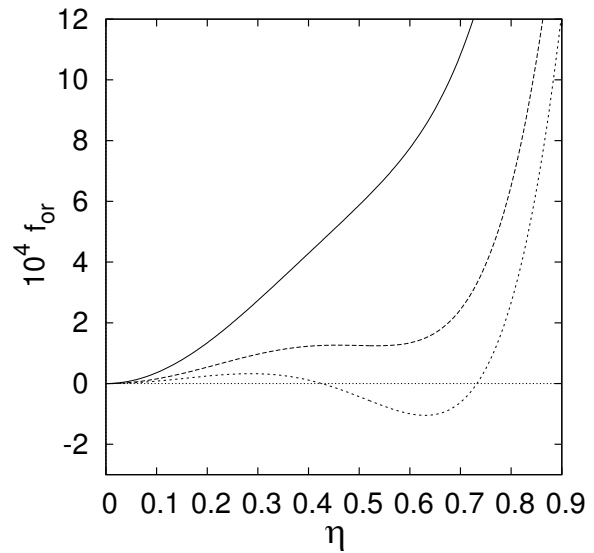}
\caption{\label{fig:PT_3fold} Plot of the orientational free energy of the 3-fold symmetric case for $L/L_p = 0.05$ and $\mu^2 = 1.1470,~1.1518,~1.1535$ (continuous, dashed, fine-dashed line, respectively).}
\end{center}
\end{figure}

\begin{figure}[h]
\begin{center}
 \includegraphics[width=8.5cm]{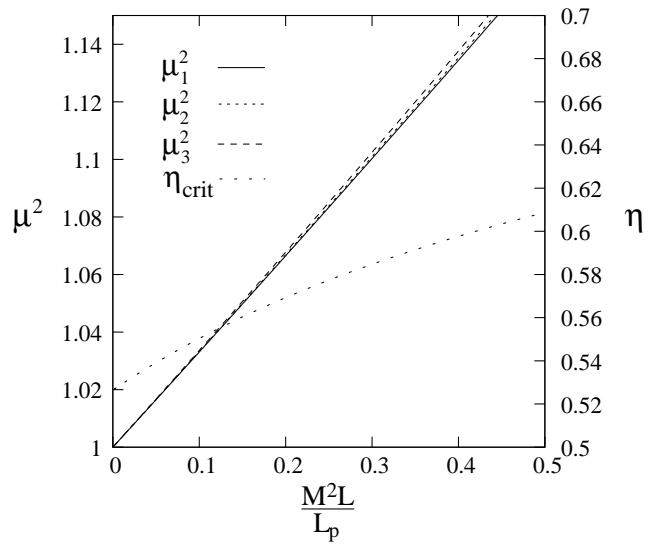}
\caption{\label{fig:m-fold_phase_trans_mus} Variational parameter of the incipient ordered state $\eta_c$ and cross-link densities $\mu_1^2$, $\mu_2^2$ and $\mu_3^2$ where a metastable ordered minimum appears, where it becomes the global minimum (phase transition) and where the minimum at zero corresponding to the disordered state becomes unstable (plot for $M>1$). Note that always $\mu_1^2 <\mu_2^2 < \mu_3^2$ for finite stiffness.}
\end{center}
\end{figure}

For finite polymer flexibility $\frac L{L_p}$ the incipient gel has no long range
orientational order. Increasing the number of cross-links beyond
$\mu^2=1$ we find a first-order transition that corresponds to a minimum of the free
energy at nonzero $\eta$, which is first metastable 
and eventually becomes the global minimum (see Fig.
\ref{fig:PT_3fold}). The critical values of $\mu^2$ for
the first appearance of a metastable ordered state ($\mu_1^2$), the first
order phase transition ($\mu_2^2$) and the disappearance of the
metastable disordered
state ($\mu_3^2$) are shown in Fig.~\ref{fig:m-fold_phase_trans_mus}.
In the limit $\frac L{L_p} \to \infty$ the phase transition coincides with the
sol-gel transition, whereas  for more
flexible filaments higher cross-link densities are required. For rather stiff polymers we find that the transition takes place at 
\begin{equation}\label{equ:crit_mu_standard}
\mu^2_2 \sim 1 + 0.33\, M^2 L/L_p \quad.
\end{equation}

In addition, Fig. \ref{fig:m-fold_phase_trans_mus} shows the
dependence of $\eta_c$, which is the value of the variational parameter $\eta$
at the phase transition, on the polymer flexibility. There is a tendency to a
lower degree of orientational localization for higher values of the
polymer stiffness. This behavior is 
qualitatively different from the behavior predicted for the lyotropic nematic
ordering of  partially flexible rods by Khokhlov and Semenov in the
corresponding range of flexibility \cite{Khokhlov_Semenov}. However, a
direct comparison of the two models is not feasible. 
In our case, orientational ordering is due to the cross-links and the critical 
density of cross-links $\mu^2_2$ of the transition {\em decreases} for increasing
polymer stiffness. There are thus two competing effects: stiffer polymers should lead to a stronger orientational localization of the polymers whereas the smaller number of cross-links should have the opposite effect.
It seems that the lower number of cross-links plays the dominant role in the dependence of $\eta_c(L/L_p)$.

We point out that the cross-linking angle $\theta = 2\pi/M$ enters the free energy as a rescaling of the polymer flexibility $L/L_p$ through the parameter $M^2$. This implies that the higher the value of $M$ (the smaller the angle), the higher the polymer stiffness required for the transition at a given cross-link density. The reason for this scaling is that within our mean-field description we are dealing with one single chain in an effective medium. M-fold order has to be propagated along the chain from one cross-link to the next and the $M^2$ scaling reflects the properties of the WLC in the calculation of the corresponding correlator.

For a first-order transition the expansion in $\eta^2$ is not really justified and can only give qualitative results. Even worse, the expansion of ${\cal F}$ in $\eta^2$ exhibits an oscillating behavior: the orders $2, 6, 10, \dots$ provide a stable orientational free energy (for large $\eta$) in a region around $\mu^2=1$ and $\frac{L}{L_p}=0$, but the orders $4,8, 12, \dots$ are always unstable in the above region because they diverge asymptotically to minus infinity. 
However, considering the purely orientational free energy
\begin{eqnarray}
f_{or, hard} &:=&\frac{\mu^2 Q^2}{2}\Bigg\{ \ln\Big( 1 + 2 \sum_{q=1}^\infty \frac{I_q^2(\eta)}{I_0^2(\eta)} \Big)\\
&-&\mu^2 \int_{s_1, s_2}\!\!\! \ln\Big( 1 + 2 \sum_{q=1}^\infty \frac{I_q^2(\eta)}{I_0^2(\eta)}  \e^{-\frac{q^2 M^2}{2 \kappa} |s_1 - s_2|} \Big) \Bigg\} \nonumber
\end{eqnarray}
it is obvious that for any finite $\kappa$ the Gaussian part is larger than the log-trace contribution as long as the cross-link density $\mu^2$ does not become too large. So, as long as we restrict ourselves to a region close enough to the sol-gel transition asymptotic stability is garantied and a qualitative picture can be obtained by truncating the expansion at order $6$, $10$ or even higher.

For the polar case, i.e $M=1$, there are additional terms ins the free energy \req{equ:var_free_energy_mfold} which couple spatial and orientational part. At first sight it might be tempting to argue that they can be neglected close to the transition because they are proportional to $\frac{1}{\xi^2} \sim Q$. As it turns out this is not correct: the orientational transition for a given polymer stiffness is shifted to significantly higher values of $\mu^2$. But all the same, the qualitative picture of the transition is still valid.

In order to analyze the orientational transition we first calculate the stationarity equation with respect to $1/\xi^2$ 
\begin{equation}
 \frac{L^2}{\xi^2}= \frac 23 \frac{\mu^2 Q}{g(\frac L{L_p}) + \eta^2\, l(\frac L{L_p}) -\frac {\eta^4} {16}\, \tilde l(\frac L{L_p}) + \frac{\eta^6}{16}\, \tilde {\tilde l}(\frac L{L_p})}
\end{equation}
and plug it again into the free energy. Keeping contributions up to order $\eta^6$ we obtain again a power series in $\eta^2$. The transition scenario that we found for $M>1$ is still valid, but the transition takes place at larger values of $\mu^2_2$. For rather stiff polymers we find approximately 
\begin{equation}
\mu^2_2 \sim 1 + 0.94 \,L/L_p
\end{equation}
and the critical $\mu^2$ is considerably larger than what we would obtain from \req{equ:crit_mu_standard} without the coupling terms.

At first sight, it seems a bit curious that the case $M=1$ is set apart by its coupling term. This is, however, only due to the low order expansion and the coupling terms for $M>1$ have not appeared \emph{yet}. Because of the higher symmetry of the orientational contributions, only higher order spatial contributions may lead to non-vanishing coupling terms.

If long range orientational order could exist only for rational values of $\theta/2\pi$, then it would be inaccessible in experiment. We show in the next section that the long range ordered states discussed above are also present if we introduce crosslinks that favor given crossing angles on average only.


\subsubsection{Soft cross-links}

Which changes do we expect when using soft
cross-links \req{equ:soft_xlink} that do not rigidly fix the intersection
angle to one particular value, but allow for fluctuations
about a given mean direction?  Soft cross-links surely will make it
harder to establish long range order, so the first expectation is that
the orientational transition will need a higher cross-link density
$\mu^2$ to take place. In particular, the behaviour in the limit of
stiff rods will change qualitatively: In the case of hard cross-links
the macroscopic network is a completely rigid object, even at
cross-link densities just above the critical value of $\mu^2 = 1$,
whereas in the case of soft cross-links there are still the degrees of
freedom of fluctuations around the preferred directions of the
cross-links left.  Instead of approaching a combined spatial \emph
{and} orientational transition right at $\mu^2 = 1$ upon increasing
the polymers stiffness, it appears possible that for soft cross-links
the long range order transition will take place not at $\mu^2=1$ but
at a higher cross-link densitiy because the networks needs to be
stabilized. However, in the limit of stiff rods \emph {and} hard
cross-links we should recover the combined transition right at $\mu^2
= 1$.

We now discuss the modifications of the free energy due to the soft
cross-links: Keeping terms only up to order $Q^3$ as before, we find
that in the case $M>1$ there is still an effective decoupling of
spatial and orientational transition. The spatial part does not change
at all, but the purely orientational contribution becomes
\begin{eqnarray}\label{equ:softx_f_or}
&&f_{or,soft}:=\\
&&\frac{\mu^2 Q^2}{2}\Bigg\{ \ln\Big( 1 + 2 \sum_{q=1}^\infty \frac{I_q^2(\eta)}{I_0^2(\eta)} \frac{I_{M q}(\gamma)}{I_0(\gamma)} \Big) \nonumber\\
&-&\mu^2 \int_{s_1, s_2}\!\!\! \ln\Big( 1 + 2 \sum_{q=1}^\infty \frac{I_q^2(\eta)}{I_0^2(\eta)} \frac{I^2_{M q}(\gamma)}{I^2_0(\gamma)} \e^{-\frac{q^2 M^2}{2 \kappa} |s_1 - s_2|} \Big) \Bigg\} \nonumber\; .
\end{eqnarray}
The soft cross-links give rise to additional factors $\frac{I_{M
    q}(\gamma)}{I_0(\gamma)}$ that equal $1$ in the limit of hard
cross-links $\gamma \rightarrow \infty$. They appear linearly in the
Gaussian and quadratically in the contribution of the log-trace
term. Their range lying between $0$ and $1$ it is clear that the
log-trace contribution is smaller with respect to the Gaussian so that
the transition occurs at a higher cross-link concentration (as
compared to the case of hard cross-links). This is confirmed by a
numerical analysis of the Taylor expansion up to 10th order in $\eta$.

\begin{figure}[h]
\begin{center}
 \includegraphics[width=8.5cm]{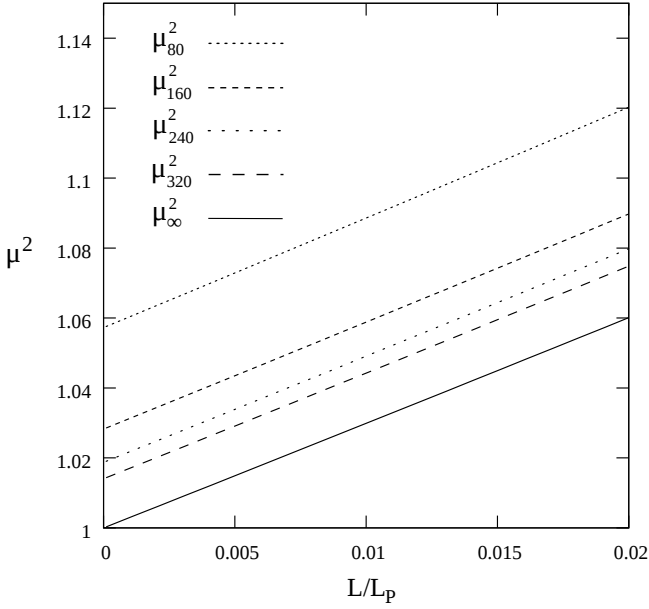}
\caption{\label{fig:3-fold_PT_softx} Plot of the critical $\mu^2(L/L_p)$ where the phase transition takes place for $M=3$ and stiffness parameters $\gamma = 80, 160, 240, 320$ and hard cross-links.} 
\end{center}
\end{figure}

The effect of soft cross-links on the orientational phase transition
is illustrated in Fig.~\ref{fig:3-fold_PT_softx} for $M=3$: First of
all, it shows that for finite $\gamma$ the phase transition takes
place at a finite distance from $\mu^2 =1$, even in the limit of stiff
rods. Moreover, comparing the curves corresponding to increasing
values of $\gamma$, i.e. to harder and harder orientational cross-link
constraints, the curves converge towards the solid curve at the bottom
that was drawn for perfectly hard cross-links as they should.

The case $M=1$ involves additional coupling terms between spatial and
orientational parameters that need to be calculated. But as before,
they don't lead to a behavior that differs qualitatively from what we
found for $M>1$.

In the preceding section we found that the higher $M$ the more
cross-links are needed to get into the long range ordered phase. This
holds true for soft cross-links, too, as we checked numerically. The
related calculations are presented in Appendices \ref{sec:m-fold_calc}
and \ref{sec:exp_val_M-fold}.


\subsection{Statistically isotropic amorphous solid (SIAS)}
In the preceding two sections, we found that (given a suitable
cross-linking angle $\theta$) there is a phase boundary
$\mu^2(\frac{L}{L_p})$ above which long range orientational order
becomes possible. For lower cross-link density, long range
orientational order vanishes. But the positional localization of the
polymer segments that takes place in the macroscopic cluster is always
accompanied by orientational localization. The corresponding
alternative to long range order are glassy states, where the average
orientations of the polymer segments are frozen in random directions,
so that isotropy is restored on a macroscopic level. We expect such
states for WLCs with a small persistence length, such that the order
induced by a cross-link cannot be sustained along the contour length
up to the next cross-link. This will be particularly severe, if $M$ is
large.

Frozen orientations for polar filaments are described by the
distribution \req{equ:ansatz_single_system_polar} for a single
site. However the direction of localization varies from chain to
chain, so that averaging over the whole sample implies an average over
the locally preferred orientation $\varphi_0$ assuming all directions
to be equally likely.  For convenience we introduce the unit vectors
${\mathbf u}^\alpha = (\cos \varphi^\alpha, \sin \varphi^\alpha)$ and
denoting the local preferential axes by ${\mathbf e}$ we get
\begin{equation}
 \int \frac{\d {\mathbf e}} {2 \pi} \;
\exp\Big(\eta {\mathbf e}\cdot (\sum_{\alpha =1}^n {\mathbf u}^\alpha) \Big)
\,=\, I_0\Big(\eta |\sum_{\alpha=1}^n {\mathbf u}^\alpha| \Big). 
\end{equation}
Altogether the order parameter for polar filaments in the
glassy state then reads
\begin{equation}
 \omega( \hat {\mathbf k}, \check \varphi ) \;=\;
\e^{-\frac{\xi^2}{2 } \hat {\mathbf k}^2} \; \frac 1 {I_0^n(\eta)} \;I_
0\Big(\eta |\sum_{\alpha=1}^n {\mathbf u}^\alpha|
\Big).\label{equ:SIAS_op_d} 
\end{equation}

What is the physical order parameter and how does it scale
with the parameter $\eta$? 

Locally, we have for each localized polymer segment a
polar moment $\left\langle {\mathbf t}_i(s)\right \rangle \sim\eta \neq 0$,
but because of the orientational disorder it vanishes globally.
We thus need an Edwards-Anderson like order parameter and choose
\begin{equation}
q=\frac{1}{N} \sum_{i=1}^N \int_s\,\left\langle {\mathbf t}_i(s)\right \rangle \cdot \left\langle {\mathbf t}_i(s)\right \rangle\;.
\end{equation}
As for the $M$-fold order parameter in the preceding section, we relate the variational parameter $\eta$ to $q$ and find that in lowest order
\begin{equation} 
q = \frac{\eta^2}4 + {\cal O}(\eta^4) \quad.
\end{equation}

We plug the Ansatz \req{equ:SIAS_op_d} for the replicated order parameter into the saddle point free energy and keep
only terms which depend on $\xi$ or $\eta$. We obtain the simplest non-trivial free energy by including terms up to second order in $\eta^2$. 
\begin{eqnarray}
\frac{{\cal{F}}}{n}& =& \frac{\mu^2 Q^2}{2} \Bigg\{- \frac{\mu^4 Q}6 \ln\frac {L^2}{\xi^2}\; + \mu^2 \frac {L^2}{4 \xi^2} g(\frac{L}{L_p})\\
&& \hspace{-5mm}-  \frac{\eta^4}{16} \Lambda(\gamma)  \bigg(1  -  \mu^2 \Lambda(\gamma) h(\frac {2L} {L_p}) \bigg) + \frac{\eta^2 L^2}{8 \xi^2} \mu^2 \Lambda(\gamma)  l(\frac{L}{L_p}) \Bigg\} \nonumber
\end{eqnarray}
where we have introduced the shorthand notation $\Lambda(\gamma) := \frac{I_1^2(\gamma)}{I_0^2(\gamma)}$.
The functions $g, h, l$ are given by \req{gofx_hofx} in the M-fold section.
Minimizing the above free energy with respect to $\xi^2$, yields qualitatively the same result as for the long range ordered state, namely
\begin{equation}
 \frac{L^2}{\xi^2}= \frac 23 \frac{\mu^2 Q}{g(\frac L{L_p}) } \quad \mbox{.}
\end{equation}

The orientational part shows a behavior different from the M-fold case: The stationarity equation with respect to $\eta^2$ gives rise to
\begin{equation}
 \eta^2 = \frac{L^2}{\xi^2} \frac{\mu^2 l(\frac L{L_p})}{1 - \mu^2 \Lambda(\gamma) h(\frac{2 L}{L_p})}
\end{equation}
The coupling terms implies a non-zero value $\eta^2$, i.e. orientational localization, as soon as positional localization sets in. Hence glassy orientational order is enslaved to positional localization and the orientational order parameter grows continuously at the gelation transition. 
Varying $\gamma$ we see that the softer the cross-links are, the smaller is also the variational parameter $\eta$, i.e. the degree of orientational order.

Note that, in contrast to the $M$-fold ordered case, the limit $n \rightarrow
0$ leads in the SIAS free energy to an extra minus sign for all the
orientational contributions because of the coupling of the replicas in the
corresponding order parameter \req{equ:SIAS_op_d}. As a
consequence, we have to maximize the free energy with respect to $\eta$
instead of minimizing it as it is well known from spin glasses \cite{spin_glass_theory}.


\subsection{Phase diagram}
\begin{figure}
\begin{center}
 \includegraphics[width=4.2cm]{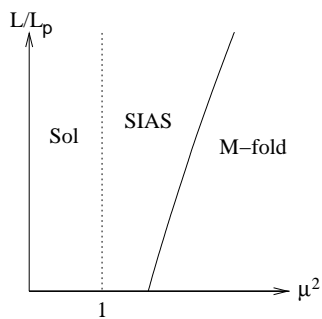}
 \caption{\label{fig:phase_diagram} Phase diagram in the plane of
   cross-link concentration $\mu^2$ and polymer flexibility $L/L_p$.
   To the right of the vertical dotted line, SIAS order becomes
   stable, wheras M-fold order appears to the right of the continuous tilted
   line which depends on M.}
\end{center}
\end{figure}
The results of the previous sections can be summarized in a phase
diagram, presented in Fig.~\ref{fig:phase_diagram}. The control
parameters are the cross-link density measured by $\mu^2$, the polymer
flexibility measured by $L/L_p$, and the cross-linking angle $\theta$. Irrespective of the
cross-linking angle, independent of the stiffness of the filaments and
of the softness of the cross-links there is a continuous gelation
transition accompanied by random local orientational ordering (SIAS
phase) at the critical cross-link density $\mu_c^2=1$. This glassy
ordering has been encountered previously for randomly linked molecules
with many legs ($p$-Beine)~\cite{p_beine,Theissen}. The free energy of
the SIAS is above the free energy of the sol, which however is unstable
beyond $\mu^2=1$ and hence, is not available in this region of the phase diagram.

What is new, is the appearance of a state with long range
orientational order, if the crossing angle $\theta
= \frac{k}{M}\, 2\pi$ where $k,M \in \mathbb Z$. This phase is characterized by a spontaneous
breaking of the rotational symmetry. For sensitive cross-links the
symmetry of the orientational order is $M$-fold, for unsensitive
cross-links and odd $M$ the resulting phase has $2M$-fold
symmetry. The free energy of the long ranged
ordered state is below the free energy of the isotropic sol as well as below the free energy of
the SIAS. Hence, we expect it to win as soon as it appears, even though
we cannot do a complete stability analysis beyond the variational
Ansatz. 

We find that the appearance of long range order is pushed to
higher values of $\mu^2$ as the constraint for the crossing angle is
softened. The same effect is observed for increasing flexibility of
the polymers, -- because it becomes more difficult to sustain the
orientation of the polymers --, and increasing $M$.

When reading the phase diagram, we should keep in mind that the parameters $\mu^2$ and $L/L_p$ are {\em not} thermodynamic ones like a temperature or a chemical potential because changing either of them changes the disorder ensemble, too. This means that two points in the phase diagram correspond effectively to two {\em different} systems.


\section{Conclusions - Discussion}
In this paper, we studied the role of the cross-linking angle in the
formation of orientational order in random networks of semiflexible
polymers in two dimensions. We have used a variational Ansatz to map
out a phase diagram. Besides a statistically isotropic amorphous solid
(SIAS) we find more exotic gels with random positional order
coexisting with long ranged orientational order, whose symmetry is
dictated by the crossing angle. In analogy to liquid crystals with
long range orientational order and thermal centre of mass motion like
in a fluid, these gel phases might be termed ``glassy crystals''-- with
long range orientational order and frozen in random positions like in
a glass. It is 
interesting to note that the tetratic ordering, which corresponds to
$M=4$ in our system, has been predicted and/or observed in a variety
of physical systems with quite different constituents and underlying
physical mechanisms \cite{Bru,Donev,Chaikin,Ramaswamy}.

Because of the peculiarities of two dimensions, we expect fluctuations to 
affect positional localization \cite{Mermin,2d_Goldbart}. It is known, in the context 
of 2d defect-mediated melting, that at finite temperature positional order 
can only be quasi-long ranged whereas orientational order can be truely 
long ranged \cite{Nelson}. A study of the corresponding 
phenomena for our positionally amorphous and orientationally ordered system is a very interesting direction for further investigation.

We point out that the finite bending rigidity is an essential
ingredient of our model. It allows the effective decoupling of
positional and orientational degrees of freedom close to the gelation
transition. In the case of infinitely stiff polymers on the other
hand, a rigid cross-link would automatically fix both position and
orientation. The nature of the emerging network is an interesting
problem which goes beyond the scope of this paper.

Another possible extension of our work concerns more elaborate variational Ans\"atze probing the appearance of combined M-fold and glassy orientational order. Here, the chains in the gel fraction are assumeed to be orientationally localized in preferential directions which vary from chain to chain but macroscopically average in an M-fold pattern.

The three-dimensional generalization of our model has to deal with the
fact that a finite cross-linking angle between two wormlike chains
prescribes a cone and not a plane. If we want to describe cross-links
with torsional rigidity, we need to go beyond the simple WLC model and use
the helical WLC \cite{Yamakawa_hwlc}.

In this work, we focused on orientational and glassy order in networks
of semiflexible polymers mediated through appropriate cross-links. We
assume that the excluded-volume interaction is such that it supports a
macroscopically translationally and rotationally invariant liquid
which, upon cross-linking, may give rise to ordered networks. Although
this assumption is mathematically consistent and facilitates the
analytical treatment of our model, it may be challenged in
experimental realizations where the excluded volume may lead to
lyotropic alignment in dense systems. In a future extension of our
model, one may envisage adding a Maier-Saupe aligning pseudopotential
as in Refs. \cite{Warner,Spakowitz} and studying its interplay with
the cross-link induced interaction.

\begin{acknowledgments}

P.B.~acknowledges support during the later part of this work by
EPSRC via the University of Cambridge TCM Programme Grant and the Project of Knowledge 
Innovation Program (PKIP) of the Chinese Academy of Sciences, Grant No. KJCX2.YW.W10.

\end{acknowledgments}


\renewcommand{\theequation}{\Alph{section}.\arabic{equation}}
\setcounter{equation}{0}  

\appendix

\section{Evaluating expectation values - the WLC propagator} \label{sec:wlc}
The wormlike chain propagator $G(\varphi, s_1; \varphi', s_2)$ quantifies the
probability that the tangential vector of monomer $s_1$ points into the
direction $\varphi$ provided that the tangential vector of monomer $s_2$ points
into the direction $\varphi'$:
\bear
&&G(\varphi, s_1; \varphi', s_2)\\
&:=& \Big\langle \delta(\varphi - \psi(s_1)) \;
\delta(\varphi' - \psi(s_2)) \Big\rangle^{{\cal H}_{WLC}} \nonumber\\
&=& \frac{1}{{\cal N}}
\int_{s_1}^{s_2} \!\! {\cal D}\{ {\mathbf t} \}\;\; \e^{- \frac
\kappa2 \int_{s_1}^{s_2} \!\d \tau \left(\frac {\d \mathbf t}{\d
\tau}\right)^2} \nonumber
\ear
Here, ${\cal N}$ denotes the normalization and $\psi(s)$ the angle corresponding to
the unity vector $\mathbf t(s)$. In principle, the path integral includes all the
monomers from $0$ to $L$, but the monomers that lie not between $s_1$ and $s_2$
do not affect the result and can be integrated out. In order to perform the
remaining path integral we write down a discretized version of the above
expression replacing the continuous degrees of freedom  by a finite number $l$ of
them such that ${\mathbf t}_1$ corresponds to ${\mathbf t}(s_1)$ and that
${\mathbf t}_l$ corresponds to ${\mathbf t}(s_2)$. The distance $\epsilon$
between neighbors is determined by $l \epsilon = | s_2 - s_1|$. Expressing the
integral and the derivatives of the wormlike chain Hamiltonian by their discretized
versions and calling the normalization constant for the discretized
path integral ${\cal N}_\epsilon$ we arrive at \cite{Kleinert}
\bear
G_{\epsilon} &=& \frac{1}{{\cal N}_\epsilon} \prod_{i=2}^{l-1} \!\!\left( \int \frac{\d \varphi_i}{2 \pi}\right) \exp\left(-\frac{\kappa}{2} \epsilon \sum_{i=1}^{l-1} \left(\frac{{\mathbf t}_i - {\mathbf t}_{i+1}}\epsilon\right)^2  \right) \nonumber \\
&=& \frac{1}{{\cal N}_\epsilon} \prod_{i=2}^{l-1} \left( \int \frac{\d \varphi_i}{2 \pi}\right)\nonumber \\
&&\hspace{1.3cm} \exp\left(-\frac{\kappa}{\epsilon}  \sum_{i=1}^{l-1} \left( 1- \cos(\varphi_i - \varphi_{i+1}) \right) \right) \nonumber 
\ear
We want now to perform the integrations over the $\varphi_i$. For that purpose it is convenient to decouple $\varphi_i$ and $\varphi_{i+1}$ by means of 
\be
\e^{a \cos \varphi} \;= \; \sum_{q=-\infty}^\infty I_q(a)\; \e^{ i q \varphi} \nonumber
\ee 
where the $I_q(a)$ denote modified Bessel functions. Performing the integrations we arrive at
\be
G_{\epsilon} = \frac{1}{{\cal N}_\epsilon} \sum_{q=-\infty}^\infty\left(\e^{-\frac{\kappa}{\epsilon}} I_q(\frac{\kappa}{\epsilon}) \right)^{l-1} \e^{i q (\varphi_1 - \varphi_l)} \nonumber 
\ee
and integrating over $\frac{\varphi_1}{2\pi}$ and $\frac{\varphi_l}{2\pi}$ we find that the normalization is given by ${\cal N}_\epsilon = \big( \exp(-\kappa / \epsilon) I_0(\kappa / \epsilon)\big)^{l-1}$. Last, we take the limit $\epsilon \rightarrow 0 $ keeping $l \epsilon = | s_1 - s_2|$ constant. It is thus possible to express the modified Bessel functions by the asymptotic expansion $I_q(a) \sim \frac{\exp(a)}{\sqrt{2 \pi a}} \big(1 - \frac{4 q^2 - 1}{8 a} + \dots \big)$ and the propagator converges to
\be \label{equ:wlc_propagator}
G(\varphi, s_1; \varphi', s_2) = \!\!\sum_{q=-\infty}^\infty \e^{- \frac{1}{2 \kappa} q^2 | s_1 - s_2| }\; \e^{i q (\varphi - \varphi')} 
\ee
Calculating a general 2-point correlation function of the (real valued) observables ${\cal O}_1$ and ${\cal O}_2$ at positions $s_1$ and $s_2$ respectively by means of the above propagator we find
\bearl{equ:propag_fourier}
&&\Big\langle {\cal O}_1(\psi(s_1))\; {\cal O}_2(\psi(s_2)) \Big\rangle^{{\cal H}_{WLC}} \\
&=& \sum_{q=-\infty}^\infty \e^{- \frac{1}{2 \kappa} q^2 | s_1 - s_2| }\; \hat {\cal O}_1(q)\; \hat {\cal O}_2^*(q ) \;\; ,\nonumber
\ear
where $\hat {\cal O}_1$ and $\hat {\cal O}_2$ denote the Fourier transformation of the observables. We learn from this expression that only the Fourier components to the same $q$ couple to each other and that the rate of the exponential decay of correlations along the filament scales with $q^2$.


\section{Mean-field replica free energy and the saddle point equations} \label{sec:F_and_SP}
In order to obtain the disorder averaged free energy $[F]$ by means of the replica method we need to calculate the disorder averaged $n$-fold replicated partition function
\begin{eqnarray}
[Z^n]&\propto& \sum_{M=0}^\infty \prod_{e=1}^M \left(\sum_{i_e,j_e=1}^N \int_{s_e, s'_e} \sum_{\sigma_e} \right) \frac{1}{M!}\left(\frac{\mu^2 V}{2N y}\right)^M \nonumber\\
&&\Big< \prod_{e=1}^M \Bigg\{ \delta(\hat {\mathbf r}_{i_e}(s_e) - \hat {\mathbf r}_{j_e}(s'_e) ) \times \\
&& \hspace{.5cm} \times \frac{\e^{\gamma \sum_{\alpha=1}^n \cos( \psi^\alpha_{i_e}(s_e) - \psi^\alpha_{j_e}(s'_e) - \theta_{\sigma_e})}}{I_0^n(\gamma)} \Bigg\} \Big>_{n+1}^{{\cal H}} \nonumber
\end{eqnarray}
where ${\mathbf r}^\alpha_i(s)$ and $\psi^\alpha_i(s)$ denote the position vector and angle of orientation of segment $s$ belonging to polymer $i$ inside the $\alpha$th replica. For sensitive cross-links $\theta_\sigma$ equals always the single crossing-angle $\theta$, i.e. the normalization $y=1$, and we can omit the summation over $\sigma$, but in the unsensitive case $\theta_\sigma$ takes the two values $\theta_1 = \theta$ and $\theta_2 = \theta + \pi$ and so, $y=2$.

Observing that the formula factorizes in the cross-link index $e$ it
is possible to perform the sum over the number of cross-links $M$ that
leads to an exponential function
\begin{eqnarray}
  [Z^n]&\propto&\Bigg< \exp\Bigg( \frac{\mu^2 V}{2N} \sum_{i,j=1}^N \int_{s,s'}  \; \delta(\hat{\mathbf r}_{i}(s) - \hat{\mathbf r}_{j}(s') ) \times\nonumber \\
  && \hspace{3cm}\times \Delta(\check \psi_{i}(s) - \check \psi_{j}(s'), \theta
  ) \Bigg) \Bigg>^{{\cal H}}_{n+1}\; \nonumber
\end{eqnarray}
where for sensitive cross-links the function $\Delta$ is defined as
\be
\Delta_s(\check \psi, \theta) \equiv \frac{\e^{\gamma \sum_{\alpha=1}^n \cos( \psi^\alpha - \theta)}}{I_0^n(\gamma)} 
\ee 
and in the unsensitive case on the other hand as
\be 
\Delta_u(\check \psi, \theta) \equiv \frac 12 \!\!\left\{ \frac{\e^{\gamma \sum_\alpha\!\! \cos( \psi^\alpha - \theta)}}{I_0^n(\gamma)} + \frac{\e^{\gamma \sum_\alpha \!\!\cos( \psi^\alpha - (\theta + \pi))}}{I_0^n(\gamma)} \right\} \;.
\ee

After the disorder average, all the sites, i.e. all the polymer
segments, are equivalent but still appear explicitly, coupled by
the cross-linking constraint $\Delta$ and the delta
functions. Expressing the delta functions in Fourier space we can
rewrite our formula in terms of the quantity
\be Q(\hat {\mathbf k}, \check m) = \frac 1{N} \sum_{i=1}^N \int_s \e^{i \hat{\mathbf k} \hat {\mathbf r}_i(s)} \; \e^{i \check m \check \psi_i(s) } \nonumber 
\ee 
Using this definition and writing ${\cal H}_{ev}$ explicitly the replicated partition function reads
\bear 
[Z^n] &\propto& \bigg\langle \exp\Big(\frac{\mu^2 N}{2 V^n} \sum_{\hat {\mathbf k}, \check m}   \Delta_{\check m} \;|Q(\hat {\mathbf k}, \check m)|^2 \Big) \nonumber\\
&&\hspace{3.5mm}\exp\Big(-\!\!\frac{\lambda^2 N^2}{2V} \sum_{\alpha=0}^n \sum_{{\mathbf k} \neq 0} \sum_m    | \rho^\alpha({\mathbf k},  m)|^2 \bigg \rangle_{n+1}^{{\cal H}_{WLC}} .\nonumber 
\ear 
In section \ref{sec:Model} the parameter $\lambda^2$ was introduced in its most general form depending on $|\mathbf k|$ and $m$, but as it turns out a constant is sufficient for our purpose and allows in the following for a more compact notation.
Because of the symmetry $|Q(\hat {\mathbf k}, \check m)|^2 = |Q(-\hat {\mathbf k}, - \check m)|^2$ it is only the real part of $\Delta_{\check m}$ that contributes and we thus redefine the sensitive kernel $\Delta_s$ accordingly as
\bel{equ:kernel_s}
\Delta_{s,\check m} = \frac{\prod_{\alpha=1}^n I_{m^\alpha}(\gamma)}{I^n_0(\gamma)}\cos\Big(\sum_\alpha m^\alpha \theta\Big).
\ee 
For unsensitive cross-links $\Delta_{u,\check m}$ equals zero if the sum $\sum_{\alpha=1}^n m^\alpha $ is not even, but takes otherwise the values of the sensitive kernel in \req{equ:kernel_s}.

We are now going to rearrange the contributions of intra-polymer repulsion and Deam-Edwards distribution into contributions belonging to the following subsets of the space of $(n+1)$-fold replicated vectors
$\hat{\mathbf k}$:
\begin{itemize}
\item The $0$-replica sector (0RS) consisting only of $\hat 0$
\item The $1$-replica sector (1RS) including all vectors of the form $\hat{\mathbf k} = (\vec 0, \dots, {\mathbf k}^\alpha,\dots,\vec 0)$ where ${\mathbf k}^\alpha \neq0$
\item The higher-replica sector (HRS) containing all the $\hat{\mathbf k}$ where wave vectors in at least to replicas are non-zero, i.e.  there are $\alpha \neq \beta \in \{0,1,\dots,n\}$ with ${\mathbf k}^\alpha\neq 0$ and $ {\mathbf k}^\beta \neq 0$ 
\end{itemize}
In the following we denote the sum over 0RS and HRS by $\overline \sum_{\hat{\mathbf k}}$ and the sum over the 1RS as $\tilde \sum_{{\mathbf k}}$. For the 1RS we obtain the new kernel 
\be
\tilde \Delta_{\check m}^\alpha = \frac {\lambda^2 N} {2V} \prod_{\beta \neq \alpha = 1 }^n \delta_{m^\beta,0}  - \frac {\mu^2}{2 V^{n}}  \; \Delta_{\check m}
\ee
Using $\tilde Q^{\alpha}({\mathbf k},\check m)$ as shorthand notation for $Q(\hat {\mathbf k},\check m)$ when only the wave vector ${\mathbf k}^\alpha$ in replica $\alpha$ is non-zero the expression reads
\bear
[Z^n] &\propto& \bigg\langle \exp\Big(- N \sum_{\alpha=0}^n  
\tilde \sum_{{\mathbf k}, \check m}   \tilde \Delta_{\check m}^\alpha \; |\tilde Q^{\alpha}({\mathbf k},\check m)|^2 \Big) \nonumber \\
&&\hspace{4mm} \exp\Big( \frac{\mu^2 N}{2 V^n} 
\overline \sum_{\hat{\mathbf k}, \check m}  \Delta_{\check m}\; |Q(\hat{\mathbf k}, \check m)|^2 \Big) \bigg \rangle_{n+1}^{{\cal H}_{WLC}} .\nonumber
\ear
In order to decouple the sites we are now going to apply a Hubbard-Stratonovich transformation. The symmetry $Q(\hat{\mathbf k}, \check m) = Q^*(-\hat{\mathbf k}, - \check m)$ and the analogous relation for $\tilde Q^\alpha({\mathbf k}, \check m)$ have to be reflected by their corresponding fields  $\Omega(\hat{\mathbf k}, \check m)$ and $\tilde \Omega^\alpha({\mathbf k}, \check m)$.
After the transformation $[Z^n]$ can be written as
\be
[Z^n] \propto \int {\cal D} \{ \tilde \Omega^\alpha, \Omega \} \exp\left(- N {\cal F}(\{ \tilde \Omega^\alpha, \Omega \}) \right) \label{equ:path_integral}
\ee 
where the integrals over the complex fields are meant to be integrations over real and imaginary parts separately. The replica free energy ${\cal F}$ is given by
\bear
{\cal F}&=& \sum_{\alpha=0}^n  \tilde \sum_{{\mathbf k}} \sum_{\check m}   |\tilde \Omega^\alpha|^2 + \frac{\mu^2 }{2 V^n} \overline \sum_{\hat{\mathbf k}} \sum_{ \check m}    |\Omega|^2  \label{equ:repl_F_prov}   \\ 
&&  - \ln \Bigg\langle \exp \bigg( 2 i \sum_{\alpha=0}^n \tilde \sum_{{\mathbf k}} \sum_{\check m} \sqrt{\tilde \Delta}\;\Re\,\Big( \tilde \Omega^\alpha \;\int_s \nonumber \\
&& \hspace{1cm}\exp\Big\{-i {\mathbf k}^\alpha {\mathbf r}^\alpha(s)\Big\} \exp\Big\{- i \check m \check \psi(s) \Big\} \Big) \nonumber \\
&&\hspace{.8cm}+\frac {\mu^2}{ V^n} 
 \overline \sum_{\hat{\mathbf k}}\sum_{\check m} \sqrt\Delta\; \Re\,\Big( \Omega  \; \int_s \nonumber\\
&&  \hspace{1cm}\exp\Big\{- i \hat{\mathbf k} \hat {\mathbf r}(s)\Big\}  \exp\Big\{- i \check m \check \psi(s) \Big\}\Big)
    \bigg) \Bigg\rangle^{{\cal H}_{WLC}}_{n+1} \nonumber
\ear
Using the saddle point approximation  we replace the integral \req{equ:path_integral} by its maximal contribution. The corresponding values of the order parameter fields have to fulfill the self-consistency equations that arise from
\bear
\frac {\partial {\cal F}}{\partial \Re \tilde \Omega^\alpha({\mathbf k}_0, \check m_0)} = 0 \;&\mbox{, }& \frac {\partial {\cal F}}{\partial \Im \tilde \Omega^\alpha({\mathbf k}_0, \check m_0)} = 0,  \nonumber \\
\frac {\partial {\cal F}}{\partial \Re \Omega(\hat{\mathbf k}_0, \check m_0)} = 0\;&\mbox{and}&\; \frac {\partial {\cal F}}{\partial \Im \Omega(\hat{\mathbf k}_0, \check m_0)} = 0. \nonumber
\ear
Note that there are many ways to perform the HS transformation, each of them leading to a different replica free energy ${\cal F}$. The resulting saddle-point replica free energy on the other hand is always the same as it can be checked easily. As a consequence, we are free to redefine the fields for later convenience by doing the replacements $\Re \Omega(\hat{\mathbf k}, \check m) \rightarrow \sqrt{\Delta_{\check m}}\;\Re \Omega(\hat{\mathbf k}, \check m)$ and for the imaginary part accordingly. The advantage of this transformation is that the saddle-point fields $\Omega_{SP}(\hat{\mathbf k}, \check m)$ are now directly related to an expectation value of $Q(\hat{\mathbf k}, \check m)$ as presented in the main part, equation  \req{equ:op_fourierspace}.
This connection between $Q$ and its corresponding field $\Omega$ is derived by means of an external field that couples to $Q$. Performing then the Hubbard-Stratonovich transformation and taking the logarithmic derivatives with respect to real and imaginary part of the field on both sides of the equation, i.e. for  both the microscopic and the field theoretic representation of our theory, results in the desired relation between $Q$ and $\Omega$.

The 0RS/HRS part of the free energy in terms of the new fields reads then
\bear
{\cal F}&=& \frac{\mu^2 }{2 V^n} \overline \sum_{\hat {\mathbf k}} \sum_{ \check m} \Delta   |\Omega|^2 \label{equ:repl_F}    \\  
&&  - \ln \Bigg\langle \exp \bigg( \frac {\mu^2}{ V^n} 
 \overline \sum_{\hat {\mathbf k}}\sum_{\check m} \Delta\; \Re\,\Big( \Omega  \; \int_s \nonumber\\
&&  \hspace{1cm}\exp\Big\{- i \hat {\mathbf k} \hat {\mathbf r}(s)\Big\}  \exp\Big\{- i \check m \check \psi(s) \Big\}\Big)
    \bigg) \Bigg\rangle^{{\cal H}_{WLC}}_{n+1} \nonumber
\ear
The 1RS part is treated in the next section.


\section{Stability of the 1-replica sector}\label{sec:1-RS}
The fields of the 1-replica sector, $\tilde \Omega^\alpha({\mathbf k}, \check m)$, can be subdivided into the fields  where $\check m$ is such that only the entry $m^\alpha$ is non-vanishing and those corresponding to the other possible values of $\check m$. In the former case the fields describe $M$-fold orientationally symmetric density fluctuations with wave vector ${\mathbf k}$ in replica $\alpha$ and have thus a clear physical meaning. The other fields break replica symmetry in the sense that they describe density fluctuations e.g. in replica $\alpha$ accompanied by purely orientational fluctuations in other replicas. These fields are unphysical and need thus to equal zero. 

In order to study the stability of the 1RS with respect to fluctuations in the physical fields $\rho^\alpha({\mathbf k}, m)$ it is sufficient to study stability for one replica only. The expansion of the corresponding free energy reads up to second order
\bear
&&{\cal F}_{1RS}= \tilde \sum_{{\mathbf k}} \sum_m |\tilde \rho({\mathbf k}, m)|^2 \\
&&\hspace{5mm}+ 2 \tilde \sum_{{\mathbf k}_1, {\mathbf k}_2} \sum_{m_1, m_2} \tilde \rho({\mathbf k}_1, m_1) \tilde \rho({\mathbf k}_2, m_2) \nonumber \\
&&\hspace{6mm} \sqrt{\tilde \Delta}_{m_1} \sqrt{\tilde \Delta}_{m_2} \underbrace{\int_{s_1, s_2} \av{\e^{i({\mathbf k}_1 {\mathbf r}_1 + {\mathbf k}_2 {\mathbf r}_2)} \e^{i (m_1 \psi_1 + m_2 \psi_2)} }}_{{\cal C}({\mathbf k}_1, {\mathbf k}_2; m_1, m_2)} \nonumber 
\ear
with $\mathbf r_x := \mathbf r(s_x)$ and $\psi_x := \psi(s_x)$.
Choosing $\lambda^2 >\mu^2$ the kernel $\tilde \Delta_m = \frac{\lambda^2}2 - \frac{\mu^2}2 \frac{I_m(\gamma)}{I_0(\gamma)}\cos(m\theta)$ is positive because $\frac{I_m(\gamma)}{I_0(\gamma)} \leq 1$. It is thus possible to redefine the fields by introducing $\overline \rho({\mathbf k}, m) := \sqrt{\Delta_m} \tilde \rho({\mathbf k}, m)$ without changing the stability. The corresponding free energy is given by
\bear
&&\overline {\cal F}_{1RS} = \tilde \sum_{{\mathbf k}} \sum_m \frac2{\lambda^2 - \mu^2 \frac{I_m(\gamma)}{I_0(\gamma)} \cos(m \theta)} |\overline \rho({\mathbf k}, m)|^2 \\
&&\hspace{1mm}+ 2 \tilde \sum_{{\mathbf k}_1, {\mathbf k}_2} \sum_{m_1, m_2}  \overline \rho({\mathbf k}_1, m_1) \overline \rho({\mathbf k}_2, m_2) {\cal C}({\mathbf k}_1, {\mathbf k}_2; m_1, m_2) \nonumber
\ear
Let us now consider the two contributions to $\overline{\cal F}_{1RS}$ separately: It is evident that the first term is a positive quadratic form provided that $\lambda^2$ is large enough. If we consider stiff rods as the limiting case of very unflexible wormlike chains the matrix of the second quadratic form can be diagonalized by switching back from the Fourier modes $m$ to the real space variables $\varphi$. Writing ${\mathbf r}(s) = {\mathbf r}_0 + s {\mathbf t}$ and integrating over ${\mathbf r}_0$ and ${\mathbf t}$ we find for ${\cal C}$
\bear
&&\int_{s_1,s_2} \Big\langle \e^{i({\mathbf k}_1 {\mathbf r}_1 +{\mathbf k}_2 {\mathbf r}_2 )} \;\delta(\varphi_1 - \psi_1) \delta\left(\varphi_2 - \psi_2\right) \Big\rangle \nonumber\\
&=& \delta_{{\mathbf k}_1, - {\mathbf k}_2}\; \delta(\varphi_1 - \varphi_2) \; \frac{\sin^2({\mathbf k}_1 {\mathbf t}_1 /2)}{\left({\mathbf k}_1 {\mathbf t}_1 /2\right)^2} 
\ear
and know thus that the second quadratic form is positive semi-definite. Altogether we have thus proved stability in the case of stiff rods, but we assume that this result will at least hold for wormlike chains with large $L/L_p$, too.


\section{Variational free energy: M-fold symmetric long ranged order \label{sec:m-fold_calc}}
In this Appendix, we present some details on the derivation of the variational free energy \req{equ:var_free_energy_mfold} of the M-fold orientationally ordered amorphous solid. We insert the corresponding replica order parameter Ansatz \req{equ:Ansatz_mfold} into the general replica free energy \req{equ:replica_F_gen_op} and expand it in powers the variational parameters $\frac{1}{\xi^2}$ and $\eta$. 

\subsection{Gaussian part} \label{sec:m-fold_calc_gaussian}
The Gaussian part reads
\bear
f_{G} &=& \frac{\mu^2 Q^2}{2 V^n} \sum_{\hat {\mathbf k}} \delta_{\sum_{\alpha=0}^n {\mathbf k}^\alpha,0} \e^{-\xi^2 \hat {\mathbf k}^2}\int_{\check \varphi_1,\check \varphi_2}\; \Delta(\check \varphi_1, \check \varphi_2)  \nonumber \\
&&  \frac{1}{I_0^{2n}(\eta)} \e^{\eta \sum_{\alpha=1}^n (\cos(M\varphi_1^\alpha) + \cos(M \varphi_2^\alpha)) } \nonumber\quad .
\ear
The spatial part is easily computed by replacing the sum over the replicated Fourier variables $\hat {\mathbf k}$ by an integral and representing the delta function in Fourier space. Performing the resulting integrations we arrive at
\be
f_{G}= \frac{1}{n+1} \left( \frac{1}{4\pi \xi^2} \right)^n f_0\;.\nonumber
\ee
For the orientational contribution $f_o$ we find
\bear \label{equ:f_or,gaussian}
&&f_{o}\\
&=&\int_{\check \varphi_1 , \check \varphi_2}\; \Delta(\check \varphi_1 - \check \varphi_2, \theta) \frac{ \e^{\eta \sum_{\alpha=1}^n (\cos(M \varphi_1^\alpha) + \cos(M \varphi_2^\alpha)) }}{I_0^{2n}(\eta)} \nonumber \\
&\sim& 1 + n \ln \left(\int_{\varphi_1, \varphi_2} \hspace{-2mm}\Delta(\varphi_1 - \varphi_2, \theta)\frac{\e^{\eta  (\cos(M \varphi_1) + \cos(M \varphi_2)) }}{I_0^{2}(\eta)}\right) \nonumber\\
&=& 1 + n \ln\Big( 1 + 2 \sum_{q=1}^\infty \frac{I_q^2(\eta)}{I_0^2(\eta)} \frac{I_{M q}(\gamma)}{I_0(\gamma)} \cos(M q \theta) \Big) \nonumber \quad .
\ear
For $\theta = \frac{k}{M} 2\pi$ the cosine equals $1$ and vanishes. In order to obtain the last line we replaced the three functions depending on the $\varphi$'s by their Fourier representations. Performing then the integrations lead to Kronecker Deltas that cancel two of the three sums over Fourier modes.

Altogether we find for the Gaussian contribution in linear order in the replica index $n$
\be
f_{G} = \frac{\mu^2 Q^2}2 \Big\{ -1 - \ln (4\pi \xi^2) + \ln\Big( 1 + 2 \sum_{q=1}^\infty \frac{I_q^2(\eta)}{I_0^2(\eta)} \frac{I_{M q}(\gamma)}{I_0(\gamma)} \Big) \Big\}. \nonumber
\ee

Considering {\bf unsensitive} crosslinker the corresponding expression reads
\begin{equation}
f_{o} = 1 + n \ln\Big( 1 + 2 \sum_{q=1}^\infty \; \delta_{M q, 2 \mathbb Z} \; \frac{I_q^2(\eta)}{I_0^2(\eta)} \frac{I_{M q}(\gamma)}{I_0(\gamma)} \cos(M q \theta) \Big)\quad ,
\end{equation}
i.e. for odd $M$ the contributions corresponding to odd $q$ are projected out. For $M$ even on the other hand, we obtain the same result as for the sensitive cross-links.

Note that for hard crosslinks ($\gamma = \infty$) there is an alternative expression for the $f_o$. For sensitive cross-links it is given by
\begin{eqnarray}\label{equ:orient_contr_gaussian_sens}
&&f_{o} \\
&\sim& 1 + n \ln \left(\int_{\varphi_1, \varphi_2} \hspace{-3mm}\Delta(\varphi_1 - \varphi_2, \theta)\frac{\e^{\eta  (\cos(M \varphi_1) + \cos(M \varphi_2)) }}{I_0^{2}(\eta)}\right) \nonumber\\
&\sim& 1 + n \ln \left( \frac{1}{I_0^{2}(\eta)} \int_{\varphi}  \e^{\eta  (\cos(M (\varphi+\theta)) + \cos(M \varphi)) }\right) \nonumber\\
&\sim& 1 + n \ln \left( \frac{1}{I_0^{2}(\eta)} \int_{\varphi}  \e^{2 \eta  (\cos(M \varphi + \frac{M \theta}2))  \cos(\frac{M \theta}{2})) }\right) \nonumber\\
&\sim& 1 + n \ln \left( \frac{I_0(2 \eta \cos(\frac{M \theta}{2}))}{I_0^{2}(\eta)} \right) \nonumber
\end{eqnarray}
and in the unsensitive case we have
\begin{eqnarray}\label{equ:orient_contr_gaussian_unsens}
 f_{o} \sim 1 + n \ln \Bigg( \frac 1 {I_0^{2}(\eta)}  \frac 12\bigg\{&&I_0\big(2 \eta \cos(\frac{M \theta}{2})\big) \\
&& + I_0\big(2 \eta \cos(\frac{M (\theta + \pi)}{2})\big)\bigg\}\Bigg) \nonumber \quad.
\end{eqnarray}


\subsection{Log-trace contributions}
Let us now turn to the log-trace contribution that we will expand up to the third order in the gel fraction $Q$: 
\bear
f_{lt} &=& \mu^2 Q f_{lt,1} + \frac{\mu^4 Q^2}2 \Big(f_{lt,2} - f_{lt,1}^2\Big) \nonumber \\
&& +\frac{\mu^6 Q^3}6 \Big( f_{lt,3} + 2 f_{lt,1}^3 - 3 f_{lt,1} f_{lt,2} \Big)  + {\cal O}(Q^4). \nonumber
\ear
The first-order term gives a trivial contribution because of
\bear
&&f_{lt,1}  \\
&=&\Bigg<\frac{1}{V^n} \; \overline \sum_{\hat {\mathbf k}} \int_{\check \varphi, \check \varphi'} \delta_{{\mathbf 0}, \sum_{\alpha=0}^n \mathbf k^\alpha}\;  \e^{-\xi^2 {\hat {\mathbf k}}^2} \times \nonumber \\
&& \hspace{0.4cm} \times \frac{\e^{\gamma\sum_{\alpha=1}^n \cos(\varphi^\alpha - \varphi'^\alpha - \theta) }}{I_0^n(\gamma)} \frac{\e^{\eta \sum_\alpha \cos(m \varphi'^\alpha)}}{{\rm I}_0^n(\eta)} \times \nonumber \\
&&\hspace{3cm} \times \int_s \; \e^{i\hat {\mathbf k} \hat {\mathbf r}(s)} \delta(\check \varphi' - \check \psi(s))\Bigg> \nonumber\\
&=& \frac{1}{V^n}\int_s \Bigg< \int_{\check \varphi} \frac{\e^{\gamma\sum_{\alpha=1}^n \cos(\varphi^\alpha - \psi(s)^\alpha - \theta) }}{I_0^n(\gamma)} \frac{\e^{\eta \sum_\alpha \cos(m \varphi^\alpha)}}{{\rm I}_0^n(\eta)} \Bigg>\nonumber \\
& =& \frac{1}{V^n} \nonumber\;.
\ear
In the second step we use $\hat {\mathbf r}(s) = \hat {\mathbf r}(0) + \int_0^s \d
\tau \;\hat {\mathbf t}(\tau)$ and transform the original path integral ${\cal D}\{\hat {\mathbf r}(s)\}$ into an integral $\frac{\d \hat {\mathbf r}(0)}{V^{n+1}}$ and a path integral ${\cal D} \{\hat \psi(s)  \}$ over angular variables. Performing the $\hat {\mathbf r}(0)$ integration we get a Kronecker delta  setting $\hat {\mathbf k}$ to zero. The last line follows from the fact that the functions are normalized and the two integrations with respect to $\check \varphi$ and $\check \psi(s)$ simply lead to $1$.\\

The second-order contribution reads
\begin{widetext}
\bear
f_{lt,2}&=&\frac{1}{V^{2n}} \int_{s_1,s_2}  \Bigg< \overline \sum_{\hat {\mathbf k}_1}  \overline \sum_{\hat {\mathbf k}_2}  \delta_{{\mathbf 0}, \sum_{\alpha=0}^n \mathbf k_1^\alpha}\; \delta_{{\mathbf 0}, \sum_{\alpha=0}^n \mathbf k_2^\alpha}\; \e^{-\xi^2 \left({\hat {\mathbf k_1}}^2 + {\hat {\mathbf k_2}}^2\right)/2}  \e^{i \left( \hat {\mathbf k_1} \hat {\mathbf r}(s_1) + \hat {\mathbf k_2} \hat {\mathbf r}(s_2) \right) }  \\
&& \hspace{1cm}\int_{ \check \varphi_1 \check \varphi_1'} \int_{ \check \varphi_2 \check \varphi_2'} \Delta\left( \check \varphi_1 -  \check \varphi_1' ,\theta \right)  \Delta \left( \check \varphi_2 - \check \varphi_2' , \theta\right)\;\delta(\check\varphi_1' - \check \psi(s_1)) \;\delta(\check\varphi_2' - \check \psi(s_2))\;\frac{\e^{\eta \sum_{\alpha=1}^n \{\cos(M\varphi^\alpha_1) + \cos(M\varphi^\alpha_2)  \} } }{{\rm I}_0^{2n}(\eta)} \Bigg>. \nonumber 
\ear
\end{widetext}	
Before Taylor expanding the expression in the variational parameters, we need to perform the summations over $\hat {\mathbf k}_1$ and $\hat {\mathbf k}_2$. We integrate over $\d \hat {\mathbf r}(0)$ as before and get a Kronecker Delta imposing $\hat {\mathbf k}_1 = - \hat {\mathbf k}_2$. There is thus only one summation left. We replace this sum by an integral and represent the Kronecker delta in Fourier space. After the two Gaussian integrations we find
\bear \label{equ:mfold_lt2}
f_{lt,2}&=&  \;\frac{1}{V^n} \left(\frac 1 {4 \pi \xi^2}\right)^n \frac 1 {n+1}  \int_{s_1,s_2} \\
&& \Bigg< \e^{\frac 1{4\xi^2} \left(\frac 1{n+1} \sum_{\alpha \beta=0}^n {\mathbf f}^\alpha {\mathbf f}^\beta - \sum_{\alpha=0}^n ({\mathbf f}^\alpha)^2\right)} \times \nonumber^\\
&& \hspace{3mm}  \times \int_{\check {\Delta \varphi_1}, \check {\Delta \varphi_2}} \frac{\e^{\gamma \sum_{\alpha=1}^n \big(\cos( \Delta \varphi_1^\alpha) + \cos( \Delta \varphi_2^\alpha) \big) }}{I_0^{2n}(\gamma)} \times  \nonumber \\
&& \hspace{3mm}\times\;\frac {\e^{\eta \sum_{\alpha=1}^n \{\cos(M (\psi_1^\alpha + \Delta \varphi_1^\alpha)) + \cos(M( \psi_2^\alpha + \Delta \varphi_2^\alpha))  \} }}{{\rm I}_0^{2n}(\eta)}  \Bigg> \nonumber
\ear
using the shorthand notations ${\mathbf f}^\alpha \equiv \int_{s_1}^{s_2} \d s \;{\mathbf t}^\alpha(s)$ and $\psi_i^\alpha := \psi^\alpha(s_i)$.
This expression can be expanded up to the desired order and the remaining task then consists in calculating the corresponding correlation functions. Details are given in Appendix \ref{sec:exp_val_M-fold}.

For the calculation of the third-order term, we proceed the same way as before and need thus to perform three sums over wave vectors. With the abbreviation ${\mathbf f}_l^\alpha \equiv \int_{s_3}^{s_l} \d \tau\; {\mathbf t}^\alpha(\tau)$ the result reads
\begin{widetext}
\bear \label{equ:mfold_lt3}
f_{lt,3}&=& \frac{1}{ V^n} \left(\frac 1 {4 \pi \xi^2}\right)^n \left(\frac 1 {3 \pi \xi^2}\right)^n \left(\frac 1 {n+1} \right)^2 \int_{s_1, s_2, s_3} \\
&&\Bigg\langle \exp\Bigg\{\frac1{3\xi^2}\bigg(\frac{1}{n+1}\sum_{\alpha, \beta=0}^n \big({\mathbf f}_1^\alpha {\mathbf f}_1^\beta + {\mathbf f}_2^\alpha {\mathbf f}_2^\beta + {\mathbf f}_1^\alpha {\mathbf f}_2^\beta\big) - \sum_{\alpha=0}^n\big( ({\mathbf f}_1^\alpha)^2 + ({\mathbf f}_2^\alpha)^2 + {\mathbf f}_1^\alpha {\mathbf f}_2^\alpha \big) \bigg)\Bigg\} \times \nonumber \\
&&  \hspace{-1cm} \times \int_{\check {\Delta \varphi_1}, \check {\Delta \varphi_2}, \check {\Delta \varphi_3}}  \frac{\e^{\gamma \sum_{\alpha=1}^n \big(\cos( \Delta \varphi_1^\alpha) + \cos(\Delta \varphi_2^\alpha) + \cos( \Delta \varphi_3^\alpha) \big) }}{I_0^{3n}(\gamma)} \; \frac {\e^{\eta \sum_{\alpha=1}^n \{\cos(M (\psi_1^\alpha + \Delta \varphi_1^\alpha)) + \cos(M( \psi_2^\alpha+ \Delta \varphi_2^\alpha)) + \cos(M (\psi_3^\alpha + \Delta \varphi_3^\alpha))  \} }}{{\rm I}_0^{3n}(\eta)}  \Bigg> \nonumber\;. \nonumber
\ear
\end{widetext}


\section{Variational free energy: SIAS}
\subsection{Gaussian part}
In the following, we present the derivation of the SIAS variational
free energy. Let us first turn to the Gaussian contribution. It is given by
\bear
f_{g} &=& \frac{\mu^2 Q^2}{2 V^n}  \sum_{\hat {\mathbf k}} \delta_{{\mathbf 0}, \sum_{\alpha=0}^n \mathbf k^\alpha} \e^{-\xi^2 {\hat {\mathbf k}}^2} \times \\
&&  \hspace{0cm} \times \int_{\check \varphi, \check \varphi'} \hspace{-4mm}\Delta(\check \varphi -  \check \varphi', \theta)\; \frac{{\rm I}_0(\eta | \sum_{\alpha=1}^n {\mathbf u}^\alpha |) {\rm I}_0(\eta | \sum_{\alpha=1}^n {\mathbf u}'^\alpha |)}{{\rm I}_0^{2n}(\eta)} \nonumber
\ear
where $\mathbf u^\alpha = (\cos \varphi^\alpha, \sin \varphi^\alpha)$.
The spatial part is treated the same way as for the M-fold case, but the orientational contribution is slightly more complicated because it does not factorize in the replica index. By means of the integral representation of the Bessel function, 
\be
I_0(\eta) = \frac 1 {2\pi} \int_0^{2\pi} \d \vartheta\; \e^{\eta \cos \vartheta}, \label{equ:IntRepresBessel}
\ee
we get a factorizing expression and find in linear order in $n$ for the orientational part of the free energy
\bear
f_{o}
&=& 1 + n \int_{\vartheta_1 , \vartheta_2 } \ln \Big( \int_{\varphi_1, \varphi_2} \Delta(\varphi_1 - \varphi_2, \theta) \times \\
&& \hspace{2.5cm} \times\frac1{I_0^2(\eta)} \e^{\eta ( \cos(\varphi_1 + \vartheta_1) + \cos(\varphi_2 + \vartheta_2)}  \Big) \nonumber \\
&=& 1+ n \int_\vartheta \ln \Bigg(1 + 2 \sum_{q \in \mathbb N} \frac{I_q^2(\eta)}{I_0^2(\eta)}  \frac{I_q(\gamma)}{I_0(\gamma)} \cos(\vartheta) \Bigg)\; . \nonumber
\ear
In order to obtain the last line, we simply switch from the angular variables to Fourier space. The expression is then easily expanded up to the desired order.

\subsection{Log-trace contributions}
The first-order contribution of the log-trace part gives the same (trivial) result as for the long range ordered case. As for the second and third order term, we again have to perform the $\hat {\mathbf k}$ summations first, but this calculation is done exactly the same way as before.The corresponding results are obtained from those of the M-fold case by simply replacing the M-fold orientational distributions in \req{equ:mfold_lt2} and \req{equ:mfold_lt3} by the corresponding SIAS distributions.

Details on the calculation of the individual terms of the second order log-trace contribution are given in Appendix \ref{sec:exp_val_SIAS}.


\section{Evaluation of the expectation values: M-fold}\label{sec:exp_val_M-fold}
In our expansion we include all the terms up to first order in $\frac 1 {\xi^2}$ and up to sixth order in $\eta$.

\subsection{Spatial part}
Let us consider the calculation of the $\frac{1}{\xi^2}$-term:
\bear
&&\int_{s_1, s_2} \left(\frac 1{n+1} \sum_{\alpha, \beta=0}^n \av{{\mathbf f}^\alpha {\mathbf f}^\beta} - \sum_{\alpha=0}^n \av{({\mathbf f}^\alpha)^2}\right)\nonumber \\
 &=& - n\;\int_{s_1,s_2} \int_{s_1}^{s_2} \d \tau \int_{s_1}^{s_2} \d \tau'\; \av{{\mathbf t}(\tau) {\mathbf t}(\tau')}\nonumber \\
&=&  -n\;\int_{s_1,s_2}  \int_{s_1}^{s_2} \d \tau \int_{s_1}^{s_2} \d \tau'\; \e^{-\frac{1}{2\kappa} |\tau-\tau'|} \nonumber \\
&=:&  -n L^2  \;g(\frac{L}{L_p}) 
\ear 
The first expectation value gives a contribution for $\alpha = \beta$ only. Combining the two terms and noticing that each replica gives the same contribution we obtain the prefactor of $-n$.
It is interesting to note that the function $g$ is closely related to the radius of gyration $R_g$. 
\begin{eqnarray}
 R_g^2 &:=& \frac{1}{2} \int_{s_1, s_2} \av{\left( \mathbf r(s_1) - \mathbf r(s_2) \right)^2} \\
&=& \frac 12 \int_{s_1, s_2} \int_{s_2}^{s_1} \d \tau \d \tau' \av{\mathbf t(\tau) \mathbf t(\tau')} \nonumber \\
&=& \frac{L^2}{2} \;g(\frac{L}{L_p}) \nonumber 
\end{eqnarray}

\subsection{Coupling terms}
In order to calculate the lowest order coupling terms we expand the spatial part in first order. For  $\theta = \frac{k}{M} 2\pi$ we find
\begin{eqnarray}\label{equ:coupl_M-fold}
 &&\frac{1}{4 \xi^2} \Bigg\langle \Big\{ \frac{1}{n+1} \sum_{\alpha \neq \beta=0}^n \mathbf f^\alpha \mathbf f^\beta - \frac{n}{n+1} \sum_{\alpha = 0}^n \left(\mathbf f^\alpha \right)^2 \Big\} \times \\
&& \hspace{1cm} \times \int_{\check {\Delta \varphi_1}, \check {\Delta \varphi_2}} \frac{\e^{\gamma \sum_{\alpha=1}^n (\cos \Delta \varphi_1^\alpha + \cos \Delta \varphi_2^\alpha)}}{I_0^{2n}(\gamma)} \times \nonumber \\
&&\hspace{1cm} \times \frac{\e^{\eta \sum_{\gamma=1}^n \left\{\cos(M(\psi_1^\gamma  + \Delta \varphi_1^\gamma))+ \cos(M(\psi_2^\gamma + \Delta \varphi_2^\gamma)) \right\} }}{I_0^{2n}(\eta)} \Bigg\rangle \;\;.\nonumber
\end{eqnarray}
We can omit the second term in the first line of the above equation because it is already proportional to $n$ and will in the end give rise to contributions  proportional to $n^2$. Writing for the moment only the terms that are directly involved in the thermal expectation value, we have
\begin{eqnarray} \label{equ:coupl_term}
 &&  \sum_{\alpha \neq \beta=1}^n \Bigg\langle \Big\{ \cos \psi^\alpha_\tau \cos \psi^\beta_{\tau'} + \sin \psi^\alpha_\tau \sin \psi^\beta_{\tau'} \Big\} \times \nonumber \\
&&\hspace{1cm}\times\e^{\eta \sum_{\gamma=1}^n \left\{\cos(M(\psi_1^\gamma  + \Delta \varphi_1^\gamma))+ \cos(M(\psi_2^\gamma + \Delta \varphi_2^\gamma)) \right\}} \Bigg\rangle	\nonumber\\
&=&-n  \;\delta_{M,1} \Bigg\{ \underbrace{\av{\cos \psi_\tau^1 \e^{\eta (c_1^1 + c_2^1)}} \av{\cos \psi_{\tau'}^2 \e^{\eta (c_1^2 + c_2^2)}}}_{(*)} \nonumber \\
&& \hspace{1.5cm}+ \av{\sin \psi_\tau^1 \e^{\eta (c_1^1 + c_2^1)}} \av{\sin \psi_{\tau'}^2 \e^{\eta (c_1^2 + c_2^2)}}\Bigg\}  \nonumber \\ 
&& \hspace{1.3cm}\times{\underbrace{\prod_{\gamma=3}^n\av{\e^{\eta (c_1^\gamma + c_2^\gamma)}}}_{(**)}}
\end{eqnarray}
where we used the abbreviations $c_i^\alpha := \cos(\psi^\alpha(s_i)+ \Delta \varphi^\alpha_i)$.
This expectation value is only non-vanishing for $M=1$ because of the relation \req{equ:propag_fourier}: the Fourier transformed of  $\cos(\psi_\tau)$ has contributions for the modes $q =\pm 1$, but the Fourier transformed of $\exp(\eta \cos(M \psi(s_1))$ only for $q = \pm \mathbb Z M$. The prefactor of $-n \sim n(n-1) $ stems from the number of ways to combine the indices $\alpha$ and $\beta$ from the spatial contribution with the $\gamma$'s from the orientational contribution.

We need thus to compute three types of expectation value. We find 
\begin{eqnarray} \label{equ:corr_func_l}
&& \frac{1}{I_0^2(\gamma)} \int_{\Delta \varphi_1, \Delta \varphi_2} \e^{\gamma \cos\Delta \varphi_1} \e^{\gamma \cos \Delta \varphi_2} \times \\
&&\hspace{7mm}\times\av{\cos(\psi_\tau) \; \e^{\eta \{ \cos(\psi_1+ \Delta \varphi_1) + \cos (\psi_2+ \Delta \varphi_2)\} }} \nonumber \\
&=& \sum_{q \in \mathbb N_0} I_q(\eta) I_{q+1}(\eta)\; \frac{I_q(\gamma) I_{q+1}(\gamma)}{I_0^2(\gamma)} \times \nonumber \\
&& \hspace{1cm}\times\frac12 \Big\{ \e^{-\frac{1}{2\kappa} ( q^2 |s_1 - \tau| + (q+1)^2 |s_2 - \tau|) } \nonumber \\
&& \hspace{1.5cm} + \e^{-\frac{1}{2\kappa} ( (q+1)^2 |s_1 - \tau| + q^2 |s_2 - \tau| )} \Big\}\;, \nonumber
\end{eqnarray}
\begin {equation}
 \av{\sin( \psi_\tau) \;\e^{\eta \{ \cos(\psi_1+ \Delta \varphi_1) + \cos (\psi_2+ \Delta \varphi_2)\} } } \;=\; 0 \;\;
\end {equation}
and
\begin{eqnarray} \label{equ:purely_orient}
 &&\frac{1}{I_0^2(\gamma)} \int_{\Delta \varphi_1, \Delta \varphi_2} \e^{\gamma \cos\Delta \varphi_1} \e^{\gamma \cos \Delta \varphi_2} \times  \\
&& \hspace{.7cm}\times\av{\e^{\eta  \{ \cos(\psi(s_1)+ \Delta \varphi_1) + \cos (\psi(s_2)+ \Delta \varphi_2)\} }} \nonumber \\
& =& I_0^2(\eta) + 2 \sum_{q \in \mathbb N} I_q^2(\eta) \; \frac{I_q^2(\gamma)}{I_0^2(\gamma)} \; \e^{-\frac{1}{2\kappa} q^2 |s_1 - s_2|}\;.\nonumber
\end{eqnarray}
 From \req{equ:coupl_term} we obtain the coefficients of the coupling terms proportional to $\frac{\eta^2}{\xi^2}$, $\frac{\eta^4}{\xi^2}$ and $\frac{\eta^6}{\xi^2}$ by sampling all the ways to collect the corresponding power in $\eta$ from the different factors. For e.g. $\frac{\eta^2}{\xi^2}$ the only way to collect an factor $\eta^2$ is to expand the two averages in $(*)$ in first order and leave out $(**)$. The result is
\begin{eqnarray}
 &&- \mu^2 \frac{1}{4 \xi^2} \Bigg\{ \eta^2  \frac{I_q^2}{I_0^2}\; l(\alpha)\\
 && \hspace{.7cm}+ \frac{\eta^4}4 \Bigg(- \frac{I_1^2}{I_0^2}\; l(\alpha) + \frac{I_1^2 I_2}{I_0^3}\; l_3(\alpha) - 4 \frac{I_1^4}{I_0^4}\; l_2(\alpha) \Bigg) \nonumber\\
&&\hspace{.7cm}+ \frac {\eta^6}4 \Bigg(\frac{155}{48} \frac{I_1^2}{I_0^2} \; l(\alpha) - \frac{5}{12} \frac{I_1^2 I_2}{I_0^3} \; l_3(\alpha) \nonumber \\ 
&&\hspace{1.7cm} + \frac{1}{16} \frac{I_1^2 I_2^2}{I_0^4} l_5(\alpha) + \frac{1}{24} \frac{I_1 I_2 I_3}{I_0^3}\; l_4(\alpha) \nonumber \\
&&\hspace{1.7cm} - \frac 14 \frac{I_1^2 I_2^2}{I_0^4} \;l_6(\alpha) - \frac{I_1^4 I_2}{I_0^5} l_7(\alpha)+ 8 \frac{I_1^4}{I_0^4}\; l_2(\alpha) \Bigg)  \Bigg\} \nonumber
\end{eqnarray}
The appearing correlation functions are defined as follows:
\begin{eqnarray}
l(x) &:=&  \frac{7 -8 \e^{- x} + \e^{-2 x} - 6 x  + 2 x^2 }{2 x^4} \\
l_2(x) &:=& \frac{-11-9 \e^{-2x} + 2 \e^{-3 x} + 18 \e^{-x } + 6 x}{9x^4}\\
l_3(x) &:=& \frac1{600 x^4} \Big(-291 - 100 \e^{-2x} + 400 \e^{-x} \\
&&\hspace{2cm}+ 16 \e^{-5x} - 25 \e^{-4x} + 180 x\Big) \nonumber\\
l_4(x) &:=& \frac 1{81000 x^4} \Big( -653- 1296 \e^{-5x}   \\
&&\hspace{0.3cm} + 2025 \e^{-4x} + 324 \e^{-10 x} - 400 \e^{-9x} + 1260x\Big) \nonumber \\
l_5(x) &:=& \frac1{7200 x^4}\Big(-297 + 400 \e^{-2 x} - 128 \e^{-5x} \\
&& \hspace{3.5cm}+ 25 \e^{-8x} + 360 x\Big) \nonumber \\
l_6(x) &:=& \frac1{1800 x^4} \Big(-37 + 100
 \e^{-6 x} - 288 \e^{-5x} \\
&&\hspace{3.3cm} + 225 \e^{-4 x} + 60x\Big) \nonumber \\
l_7(x) &:=& \frac1{1350 x^4} \Big(-114 - 100 \e^{-3x} + 225 \e^{-2x}\\
&&\hspace{3cm} + 25 \e^{-6 x} - 36 \e^{-5x} + 120 x \Big) \nonumber
\end{eqnarray}
In the case of {\bf hard cross-links} it is convenient to introduce the following functions:
\begin{equation}
 \tilde l(x) := 16 \Big( l(x) + 4\, l_2(x) -l_3(x) \Big)
\end{equation}
and
\begin{eqnarray}
 \tilde {\tilde l}(x) &:=& \Big( \frac{155}{3} l(x) + 128\, l_2(x) - \frac{20}{3} l_3(x) \\
&&+ \frac 23 l_4(x) + l_5(x) - 4\, l_6(x) - 16\, l_7(x) \Big) \nonumber
\end{eqnarray}

\subsection{Orientational contributions}
As for the calculation of the purely orientational part we notice first that it is factorizing in the replica index $\alpha$. It is thus possible to take the replica limit directly by means of the expansion $ \av{\dots}^n = 1 + n \ln \av{\dots} + {\cal O}(n^2)$ and we have
\begin{eqnarray}
f_{or} &=& 1 + n \int_{s_1, s_2} \hspace{-.5cm}\ln \Bigg(\Big\langle \int_{ \Delta \varphi_1, \Delta \varphi_2} \hspace{-5mm}\frac{\e^{\gamma (\cos \Delta \varphi_1 + \cos \Delta \varphi_2)}}{I_0^2(\gamma)} \times\\
&& \hspace{1.5cm}\times \frac{\e^{\eta (\cos(M(\psi_1+ \Delta \varphi_1))+ \cos(M(\psi_2 + \Delta \varphi_2))}}{I_0^2(\eta)}  \Big\rangle \Bigg)\nonumber
\end{eqnarray}
The thermal average can be performed using formula \req{equ:propag_fourier}. Keeping only the contribution linear in the replica index we obtain
\begin{eqnarray}
f_{or} &=& \int_{s_1, s_2} \hspace{-3mm}\ln \Bigg(\Big\langle \int_{ \Delta \varphi_1, \Delta \varphi_2} \hspace{-5mm}\frac{\e^{\gamma (\cos \Delta \varphi_1 + \cos \Delta \varphi_2)}}{I_0^2(\gamma)} \times\\
&& \hspace{7mm} \times \sum_{q \in \mathbb Z} \e^{-\frac{q^2}{2\kappa} |s_1 - s_2|} \frac{I_q^2(\eta)}{I_0^2(\eta)} \e^{i q (\Delta \varphi_1 - \Delta \varphi_2)}\Big\rangle \Bigg)\;.\nonumber
\end{eqnarray}
The $\Delta \varphi$-integrations lead to a Fourier transformation of the soft cross-link contributions and noticing that the resulting expression is symmetric in $q$ we find
\begin{eqnarray}\label{equ:orient_contr_logtrace}
f_{or} = \!\!\!\int_{s_1, s_2} \hspace{-5mm} \ln\Bigg( \!1 \!+\! 2 \!\!\sum_{q=1}^\infty \frac{I_q^2(\eta)}{I_0^2(\eta)} \frac{I^2_{M q}(\gamma)}{I^2_0(\gamma)} \e^{-\frac{q^2 M^2}{2 \kappa} |s_1 - s_2|} \!\Bigg)
\end{eqnarray}	
Expanding the orientational free energy for {\bf hard cross-links} in $\eta$ and performing the $s_1$- and $s_2$-integrations it is convenient to introduce the function $h(x)$ 
\be
h(\frac{L}{L_p}) \equiv \int_{s_1, s_2} \e^{-\frac{1}{2 \kappa}  |s_1 - s_2|} \label{equ:h(alpha)}
\ee
and to build of $h(x)$ the functions
\be
\tilde h(x) \equiv h(x) + h(2x) - \frac 14 h(4x)
\ee
and 
\bear
\tilde {\tilde h}(x) &\equiv& \frac{11}{16} h(x) +3 h(2 x ) + h(3 x) \\
&& - \frac 14 h(4 x) - \frac 38 h(5 x) + \frac 1{48} h(9 x)\;. \nonumber
\ear


\section{Evaluation of the expectation values: SIAS} \label{sec:exp_val_SIAS}
The calculation of the expectation values in the second order contributions of the SIAS log-trace part can be done along the same lines as above using again the integral representation of the Bessel function \req{equ:IntRepresBessel}. The calculation of the lowest order spatial contributions is the same as for the M-fold case. For the {\bf purely orientational} we obtain
\begin{eqnarray}
f_{lt2, or} &=& \int_{s_1, s_2} \int_{\vartheta_1, \vartheta_2} \ln\Big( 1 + 2 \sum_{q=1}^\infty \frac{I_q^2(\eta)}{I_0^2(\eta)} \frac{I^2_{q}(\gamma)}{I^2_0(\gamma)} \times \\
&& \hspace{2cm}\times \e^{-\frac{q^2}{2 \kappa} |s_1 - s_2|} \cos(q (\vartheta_1 - \vartheta_2)) \Big) \Bigg\}\; . \nonumber
\end{eqnarray}
The lowest order {\bf coupling term} is given by
\begin{eqnarray}
&& \hspace{-3mm}\frac{1}{4 \xi^2} \Bigg\langle \!\!\!\left( \frac{1}{n+1}\!\! \sum_{\alpha \neq \beta =0}^n \mathbf f^\alpha \mathbf f^\beta  \!-\! \frac{n}{n+1}\sum_{\alpha=0}^n (\mathbf f^\alpha)^2  \right) \times \\
&& \hspace{7mm} \times \int_{\vartheta_1, \vartheta_2} \int_{\check {\Delta \varphi_1}, \check {\Delta \varphi_2}}\hspace{-4mm} \frac{\e^{\gamma \sum_{\alpha=1}^n (\cos \Delta \varphi_1^\alpha + \cos \Delta \varphi_2^\alpha)}}{I_0^{2n}(\gamma)} \times \nonumber \\
&& \hspace{-3mm}\times \frac{\eta^2}{2} \!\!\!\sum_{\gamma, \delta=1}^n\!\!\! \Big(\!\!\cos(\psi_1^\gamma\! - \!\vartheta_1 \!+ \!\Delta \varphi_1^\gamma)\! +\! \cos(\psi_2^\gamma \!-\! \vartheta_2 \!+\! \Delta \varphi_2^\gamma) \Big) \nonumber \\
&&\hspace{-3mm}\times \Big(\cos(\psi_1^\delta - \vartheta_1 + \Delta \varphi_1^\delta) + \cos(\psi_2^\delta - \vartheta_2 + \Delta \varphi_2^\delta) \Big) \Bigg\rangle \nonumber
\end{eqnarray}
As in \req{equ:coupl_M-fold} the second term in the first line of the expression is already proportional to $n$ and will thus finally lead to a contribution of ${\cal O}(n^2)$ that is not of interest for us. After integrating with respect to the $\vartheta$-variables and then with respect to the $\varphi$-variables we obtain in the end
\begin{equation}
 -n \frac{\eta^2}{8 \xi^2} \frac{I_1^2(\gamma)}{I_0^2(\gamma)} l(\alpha)\qquad \mbox{.}
\end{equation}
The results for the SIAS can be expressed in terms of the functions $g$, $h$ and $l$ that we already introduced in the preceding section.


\bibliographystyle{apsrev}
\bibliography{martin}

\end{document}